\documentclass[10pt,final,journal]{IEEEtran}
\usepackage{graphicx}
\usepackage{stfloats}
\usepackage{multirow} 
\usepackage{multicol} 
\usepackage{xcolor}
\usepackage{amsmath}
\usepackage{amssymb}
\usepackage{array}
\usepackage{upgreek}
\usepackage[caption=false,font=footnotesize]{subfig}
\usepackage[numbers,sort&compress]{natbib}
\usepackage{url}
\usepackage{booktabs}
\usepackage{bbm}
\usepackage[bookmarks,colorlinks]{hyperref} 

\usepackage[boxed,ruled,linesnumbered,commentsnumbered]{algorithm2e}

\usepackage{clipboard}
\newclipboard{myclipboard}

\ifCLASSINFOpdf
\else
\fi

\begin{document}
%
\title{Decentralized Automotive Radar Spectrum Allocation to Avoid Mutual Interference \\ Using Reinforcement Learning}
%
%
%

\author{Pengfei Liu,
        Yimin~Liu\textsuperscript{*}, ~\IEEEmembership{Member,~IEEE}, ~Tianyao~Huang,
         ~Yuxiang~Lu, and ~Xiqin~Wang
\thanks{P. Liu, Y. Liu, T. Huang and X. Wang are with the Department of Electronic Engineering, Tsinghua University, Beijing, China (e-mail: {liupf16@mails.tsinghua.edu.cn}; {yiminliu@tsinghua.edu.cn}; {huangtianyao@tsinghua.edu.cn}; {wangxq\_ee@tsinghua.edu.cn}).}
\thanks{Y. Lu is with Beijing Radar Research Institute, Beijing, China (email: {luyuxiang92@163.com}).}
\thanks{\textsuperscript{*}Corresponding author.}
}

\maketitle
\begin{abstract}
Nowadays, mutual interference among automotive radars has become a problem of wide concern. In this paper, a decentralized spectrum allocation approach is presented to avoid mutual interference among automotive radars. Although decentralized spectrum allocation has been extensively studied in cognitive radio sensor networks, two challenges are observed for automotive sensors using radar. First, the allocation approach should be dynamic as all radars are mounted on moving vehicles. Second, each radar does not communicate with the others so it has quite limited information. A machine learning technique, reinforcement learning, is utilized because it can learn a decision making policy in an unknown dynamic environment. As a single radar observation is incomplete, a long short-term memory recurrent network is used to aggregate radar observations through time so that each radar can learn to choose a frequency subband by combining both the present and past observations. Simulation experiments are conducted to compare the proposed approach with other common spectrum allocation methods such as the random and myopic policy, indicating that our approach outperforms the others.
\end{abstract}


%
\IEEEpeerreviewmaketitle

\section{Introduction}
%
%
%
%

The pursuit of safe and comfortable driving has recently given rise to the advance of self-driving technologies, such as adaptive cruise control and collision warning.  Automotive radar, as one of the most important sensors on vehicles, is vastly popularized. In most countries, the frequency range of 76-77 GHz is allocated to automotive usage \cite{Goppelt2010Automotive}. As the population and bandwidth demand of automotive radars are both on the rise, mutual interference becomes a problem of wide concern. 

Frequency modulated continuous wave (FMCW) is widely employed in automotive radar due to its low hardware complexity \cite{kim2018peer}. The consequences of mutual interference among FMCW radars have been investigated in \cite{Brooker2007Mutual, Goppelt2010Automotive,goppelt2011analytical,Luo2013A,toth2018analytical}. The probability of ghost targets is raised if an FMCW radar is interfered by another with the same chirp rate \cite{Brooker2007Mutual, Goppelt2010Automotive,goppelt2011analytical}. To prevent ghost targets, it is proposed in \cite{Luo2013A} to use random chirp rates so that interference causes a rise in the noise level. Such interference is referred to as non-coherent interference in \cite{toth2018analytical}. The MOSARIM project also concluded that the most probable consequence of real-world automotive radar interference was an increase in receiver noise which might cover targets \cite{kunert2010final}.

Current solutions to mitigating automotive radar interference can be grouped into two categories, interference canceling and interference avoidance. \textcolor{black}{Interference canceling techniques are usually applied in the radar receiver to suppress interference in time \cite{bechter2015automotive}, frequency \cite{wagner2018threshold} and time-frequency domain  \cite{barjenbruch2015method}. For example, in \cite{bechter2015automotive}, the interference signal is first reconstructed in time domain by parameter estimation and then subtracted from the received signal. In \cite{wagner2018threshold}, a minimum  operation is performed on a set of chirps in the frequency domain to suppress noise-like interference. In \cite{barjenbruch2015method}, an algorithm is proposed to locate and then eliminate the samples contaminated by interference in time-frequency domain.}
Interference avoidance is to coordinate transmission in time, frequency and space domain to prevent interference from occurring \cite{alland2019interference}. A representative interference avoidance method in the frequency domain is spectrum allocation. For example, in \cite{kunert2010study}, an allocation scheme is described in which the whole band is equally divided into several non-overlapping subbands. The bandwidth of each subband is determined by the resolution requirement. Then, radars are assigned to different subbands so they do not interfere with each other. However, the interference is inevitable when radars outnumber subbands. In \cite{Khoury2016RadarMAC}, a centralized spectrum allocation approach is proposed. Each radar sends information including its own position and velocity to a control center, which computes the allocation results and then broadcasts them to each radar. However, it increases extra communication cost. By contrast, in decentralized allocation, each radar chooses their frequency subbands in an autonomous way. A straightforward method is to choose at random \cite{Khoury2016RadarMAC}, which is easy to implement but the mitigation of interference is limited. 

In this paper, we present an interference avoidance approach for FMCW automotive radar by decentralized spectrum allocation. In our approach, like \cite{kunert2010study}, we also assume that the whole band is equally divided into several non-overlapping subbands, given the resolution or bandwidth requirement. Moreover, we consider the cases where radars outnumber subbands. Based on the premise above, we propose a decentralized spectrum allocation approach in which each radar chooses a subband separately to reduce the mutual interference. 
Although decentralized spectrum allocation has been extensively studied in cognitive radio sensor networks, two challenges are observed for automotive sensors using radar. First, the allocation approach should be dynamic as all radars are mounted on moving vehicles. Second, each radar does not communicate with the others so it has quite limited information. In light of these challenges, a machine learning technique, reinforcement learning (RL), is employed on each radar since RL can learn a decision making policy in an unknown dynamic environment. Moreover, a long short-term memory (LSTM) recurrent network is utilized to aggregate  observations through time so that radar can learn to choose a subband by combining both the present and past observations.



\textcolor{black}{
The proposed approach to solve automotive radar interference problem, in a broad sense, lies in the realm of spectrum sharing, which was first proposed in the field of communication to accommodate 
more radio frequency services and avoid mutual interference. More recently, spectrum sharing techniques have been explored extensively in radars so that they can share the scarce spectrum with both communication and other radar systems. In these techniques, two categories are reviewed in this paper. One is the co-design of the waveform for both radar and communication \cite{ bliss2019communications}. In \cite{ bliss2019communications}, several metrics are considered to evaluate the shared waveform, such as spectral efficiency for communication and estimation performance for radar. The other category is the coexistence \cite{cohen2017spectrum,martone2017spectrum},  in which radar first uses spectrum sensing to obtain the occupancy of the whole band and then chooses proper subbands by solving an optimization problem. In \cite{cohen2017spectrum}, considering that a wide band needs to be sensed, a compressed sampling technique is proposed to reduce the sampling and processing requirements. In \cite{martone2017spectrum}, because the  optimization to improve the signal-to-interference plus noise ratio is at high computational cost, a bioinspired filtering technique is proposed to reduce the computational complexity. Although we also present an approach to achieve the coexistence of multiple automotive radars, our work differs from \cite{cohen2017spectrum,martone2017spectrum} in that an extra spectrum sensing receiver is not required. Radar only needs its own receiver to estimate the interference power within the subband on which it transmits.}

\textcolor{black}{With the proposing of cognitive radar \cite{Haykin2006Cognitive},  in some researches such as \cite{griffiths2014radar,greco2018cognitive},  it is indicated that cognitive approaches can be used to reduce the interference between radar and other radio frequency systems. In terms of spectrum sharing, radar cognition includes observation of the spectral environment and decision-making for the transmission \cite{greco2018cognitive}. In recent years,  RL, which is a machine learning method for decision-making, has been adopted in radar spectrum sharing problems. For example, in \cite{kozy2019applying}, deep Q-learning, which is an RL approach combined with deep neural networks, is utilized so that radar learns to choose subbands to avoid the interference from a communication system. Like \cite{cohen2017spectrum,martone2017spectrum}, in \cite{kozy2019applying}, radar also needs to observe the occupancy of all subbands while deciding the transmitting subband. Whereas, in this paper, each radar only observes the subband on which it transmits by estimating the received interference power. Moreover, recurrent neural networks are adopted to aggregate radar observations through time so that radar can learn to choose a subband by combining the current and past observations. However, in \cite{kozy2019applying}, a fully connected network is used and radar chooses a subband based on only a single observation.
} 

Beside the fields of spectrum sharing, RL has also been successfully applied to other cognitive radar applications, such as cognitive electronic warfare  \cite{kang2018reinforcement,you2019deep} and waveform optimization \cite{wang2018reinforcement}. However, only a single radar is considered in these researches. In this work, we investigate a multi-radar interference avoiding problem in which each automotive radar cognitively changes subbands according to their observations.
%

In communication, the multi-user interference problem has also been investigated using RL. In \cite{li2010multiagent,faganello2013improving,wang2018deep,chang2018distributive,naparstek2018deep}, a dynamic channel/subband access policy is learned so that each user can avoid colliding into the same channel/subband with others. In these researches, users are assumed to be static. Whereas, in our problem, radars are mounted on moving vehicles, so the radio environment changes with the positions of radars. To cope with such situation, we first address how radar acquires the position-related observations, which include not only the detected distance of other radars, but also the estimated interference power, which decreases with the distance. Then, we show how radar exploits these observations to choose a proper subband.

To sum up, the main contributions of this paper are listed as follows.
\begin{itemize}
	\item A decentralized spectrum allocation approach for automotive radar is proposed using RL. Each radar learns to choose a subband to avoid interference according to its own observations, with almost no communication required. 
	
	\item The LSTM network is utilized in the RL-based spectrum allocation approach so that radar can learn to choose a subband by combining its current and past observations. Moreover, an algorithm to train the LSTM network in our problem is presented.


	\item Simulation experiments are conducted to compare the proposed approach with other common decentralized spectrum allocation methods such as the random and myopic policy, showing that our approach outperforms the others.

\end{itemize}

The rest of the paper is organized as follows. In Section II, the scenario and signal model are constructed. In Section III, how radar measures the range, velocity and interference is explained. In Section IV, the decentralized spectrum allocation approach using RL with LSTM networks is elaborated. Simulation results and concluding remarks are presented in Section V and VI, respectively.

\section{Scenario and Signal Model}
\subsection{Scenario Model}
\begin{figure}[!t]
	\centering
	\includegraphics[width=0.5\textwidth]{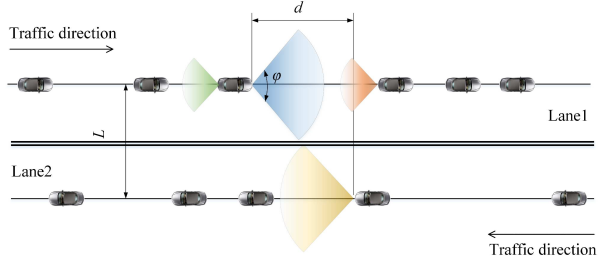}
	\caption{Problem scenario.}
	\label{fig:scenario}
\end{figure}

A simplified scenario considered in our problem is shown in Fig. \ref{fig:scenario}, in which cars are traveling in two lanes with different traffic directions. Each car is equipped with one long-range radar on its front and one short range radar on its back. The long-range radar is used to provide a forward-looking view for applications such as adaptive cruise control and collision mitigation systems \cite{Hasch2012Millimeter}. The short-range radar is used to detect obstacles for applications such as lane change assistance and assisted parking systems \cite{Hasch2012Millimeter}.

As is mentioned in the previous section, it is presumed that the whole  band is equally divided into $M$ non-overlapping subbands. Suppose there are $N$ cars on a certain section of the road and they are indexed by $1,2,...,N$. In each period, a car chooses a working subband for both the long-range and short-range radar. The length of a period is $T$. As radars outnumber subbands, i.e. $N>M$, more than one radars will inevitably collide into the same subband causing interference. In this paper, we only focus on reducing LRR interference, as SRR interference is usually not a concern \cite{Luo2013A}. 

\subsection{Transmitted Signal Model}
\label{subsec:sig_model}
\begin{figure}[!t]
	\centering
	\includegraphics[width=0.5\textwidth]{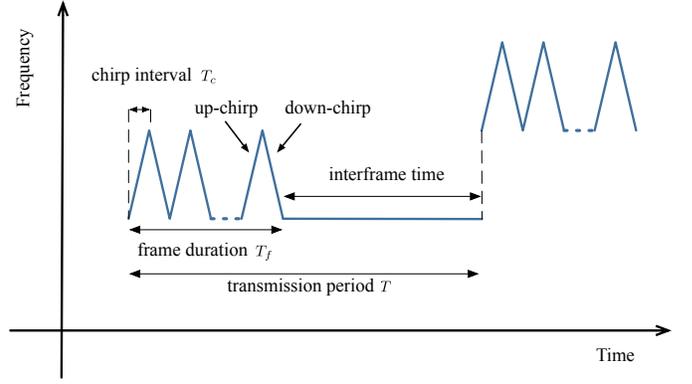}
	\caption{\textcolor{black}{Illustration of the triangular chirp FMCW waveform used in this paper. In this figure, radar chooses different subbands for the two transmission periods.}}
	\label{fig:waveform}
\end{figure}

In this work, the triangular chirp FMCW waveform \cite{Luo2013A, Khoury2016RadarMAC} is adopted. The transmitted waveform is 
\begin{equation}
s(t)=\left\{\begin{array}{l}
\exp\left(j\pi\frac{B}{T_c}t^2\right)\exp\left(j2\pi f_mt\right)\\
\qquad\qquad\qquad\qquad\qquad\qquad 0\le t<T_c\\
\exp\left(-j\pi\frac{B}{T_c}(t-2T_c)^2\right)\exp\left(j2\pi f_mt\right)\\
\qquad\qquad\qquad\qquad\qquad\qquad T_c \le t < 2T_c \end{array}
\right.,
\label{trans_signal}
\end{equation} 
where $T_c$ is the chirp interval, $B$ is the chirp bandwidth, and $f_m=f_0+mB$ is the carrier frequency for the $m$th subband. The chirp interval $T_c$ determines the chirp rate $\dfrac{B}{2T_c}$, since the bandwidth $B$ is constant. \textcolor{black}{For convenience, in (\ref{trans_signal}), we only express the waveform in two chirp intervals. One is the up-chirp whose instantaneous frequency increases with time. The other is the down-chirp whose instantaneous frequency decreases with time. The triangular chirp FMCW waveform  used in this paper is illustrated in Fig. \ref{fig:waveform}. For each transmission, radar transmits a train of triangular chirps, which is called a chirp frame. The frame duration is $T_f$. For each transmission, radar can choose a different subband. The transmission period is $T$.}

\textcolor{black}{In our problem, parameters including the bandwidth $B$, frame duration $T_f$ and transmission period $T$ are fixed for all radars. Moreover, radars on different cars use a different chirp interval $T_c$ \footnote{In this paper, the chirp number $N$ is determined by $N=\lfloor T_f/T_c \rfloor$, where $\lfloor\cdot\rfloor$ means the nearest integer less than or equal to that element.Actually, in real FMCW radars, the frame duration is also slightly larger than the chirp duration times the chirp number considering the short idle time between adjacent chirps \cite{dham@2017programming}.}. This practice is to avoid ghost targets, which will be elaborated in \ref{subsec:interference}. The purpose of our work is to enable each radar to learn to choose a subband for each transmission so that mutual interference can be avoided.}

\subsection{Received Signal Model}
The received signal is composed of the target echo, interference and noise:
\begin{equation}
r(t) = e(t)+h(t)+n(t),
\label{received_signal}
\end{equation}
where $e(t)$, $h(t)$ and $n(t)$ denote the echo, interference and noise, respectively.

The echo from a single target is a delayed version of the transmitted signal:
\begin{equation}
e(t) = \sqrt{P_S}\cdot s\left(t-\tau(t)\right),
\label{echo}
\end{equation}
where $P_S$ is the received signal power and $\tau(t)$ the time delay of the target reflection. For an approaching target with relative radial velocity $v$,
\begin{equation}
\tau(t) = \frac{2(D-vt)}{c},
\label{target_delay}
\end{equation}
where $D$ is the target radial distance, $v$ is the target relative radial velocity, and $c$ is the light speed. 

Likewise, the received interference signal from another radar with transmitted signal $s'(t)$ is 
\begin{equation}
h(t) = \sqrt{P_I}\cdot s'\left(t-\tau'(t)\right),
\label{interference}
\end{equation}
where $P_I$ is the received interference power and $\tau'(t)$ the time delay:
\begin{equation}
\tau'(t) = \frac{(D'-v't)}{c},
\label{interference_delay}
\end{equation}
where $D'$ is the the radial distance and $v'$ is the relative radial velocity of the interfering radar. 

\begin{figure}[!t]
	\centering
	\includegraphics[width=0.4355\textwidth]{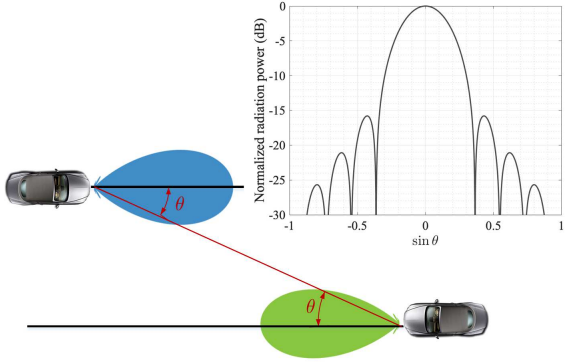}
	\caption{Illustration of radar antenna pattern.}
	\label{fig:antenna_pattern}
\end{figure}

The received interference power $P_I$ depends on the relative positions of the two radars. If they are located on different lanes (indicating the interfering radar is a long-range radar),
\begin{equation}
P_I = \frac{P_LGA_eg}{4\pi\left(L^2+d^2\right)} \cdot \left[p_r\left(\theta\left(d\right)\right)\right]^2,
\label{interference_LRR}
\end{equation}
where 
\begin{itemize}
	\item $P_L$: transmitting power of the long-range radar;
	\item $G$: antenna gain;
	\item $A_e$: effective area;
	\item $L$: vertical distance between two lanes;
	\item $d$: horizontal distance between two radars;
	\item $\theta$: radiation direction between two radars;
	\item $p_r(\cdot)$: normalized antenna beam pattern;
	\item $g$: propagation decaying factor.
\end{itemize}
In (\ref{interference_LRR}), the antenna pattern $p_r(\cdot)$ is taken into consideration, which indicates that the transmitting or receiving power also depends on the direction from one radar to another. An illustration of the antenna pattern is provided in Fig. \ref{fig:antenna_pattern}. The direction can be written as a function of $d$:
\begin{equation}
\theta(d) = \arctan\left(\frac{L}{d}\right).
\end{equation}

If the two radars are located on the same lane (indicating the interfering radar is a short range radar), 
\begin{equation}
P_I=\frac{P_SGA_eg}{4\pi d^2},
\label{interference_SRR}
\end{equation}
where $P_S$ is the transmitting power of the short-range radar. The antenna pattern is omitted here because the direction is approximately $0$.


\begin{figure*}[!t]
	\centering
	\subfloat[No interference]{\includegraphics[width=0.33\textwidth]{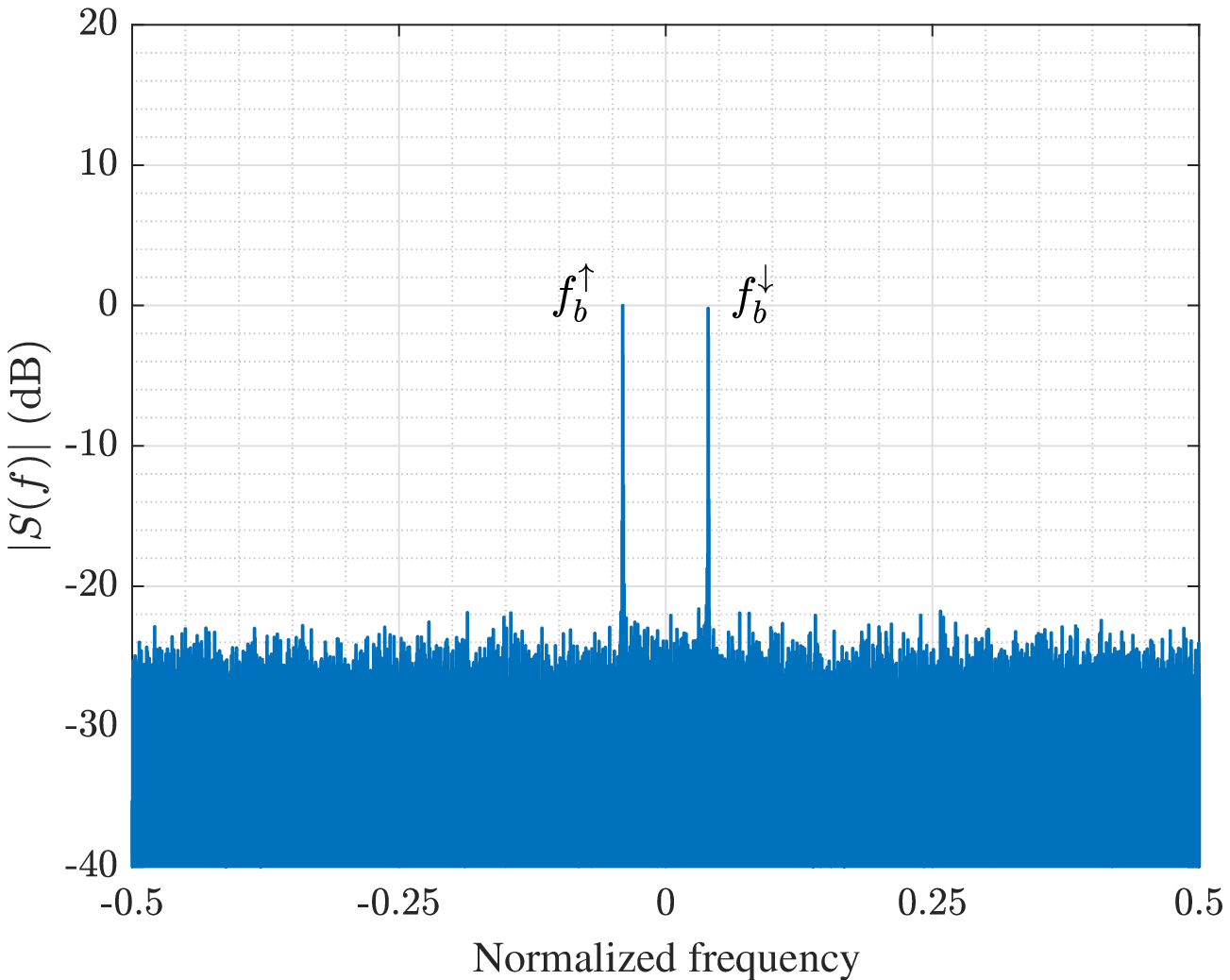}}
	\subfloat[Ghost targets]{\includegraphics[width=0.33\textwidth]{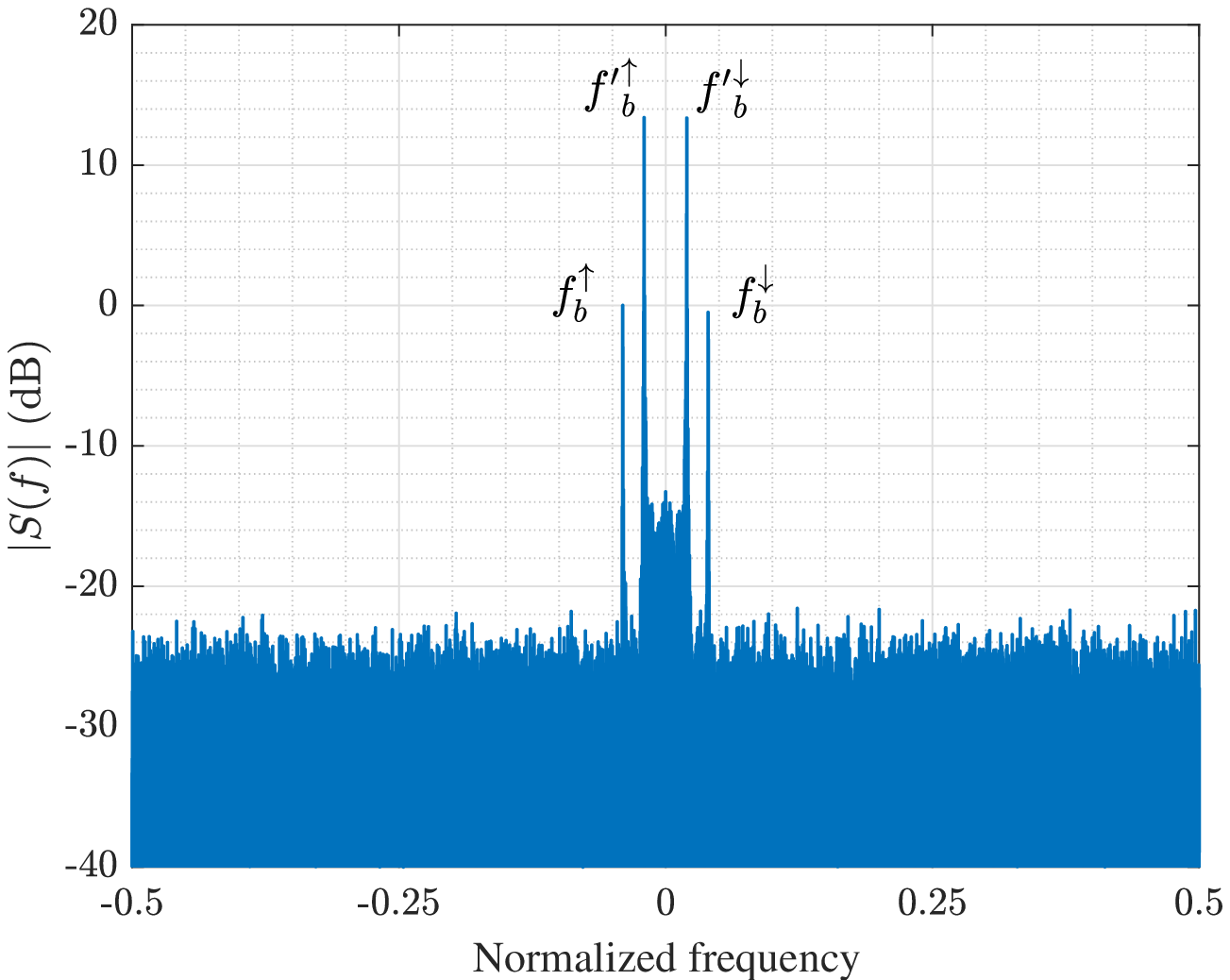}}
	\subfloat[Raised noise level]{\includegraphics[width=0.33\textwidth]{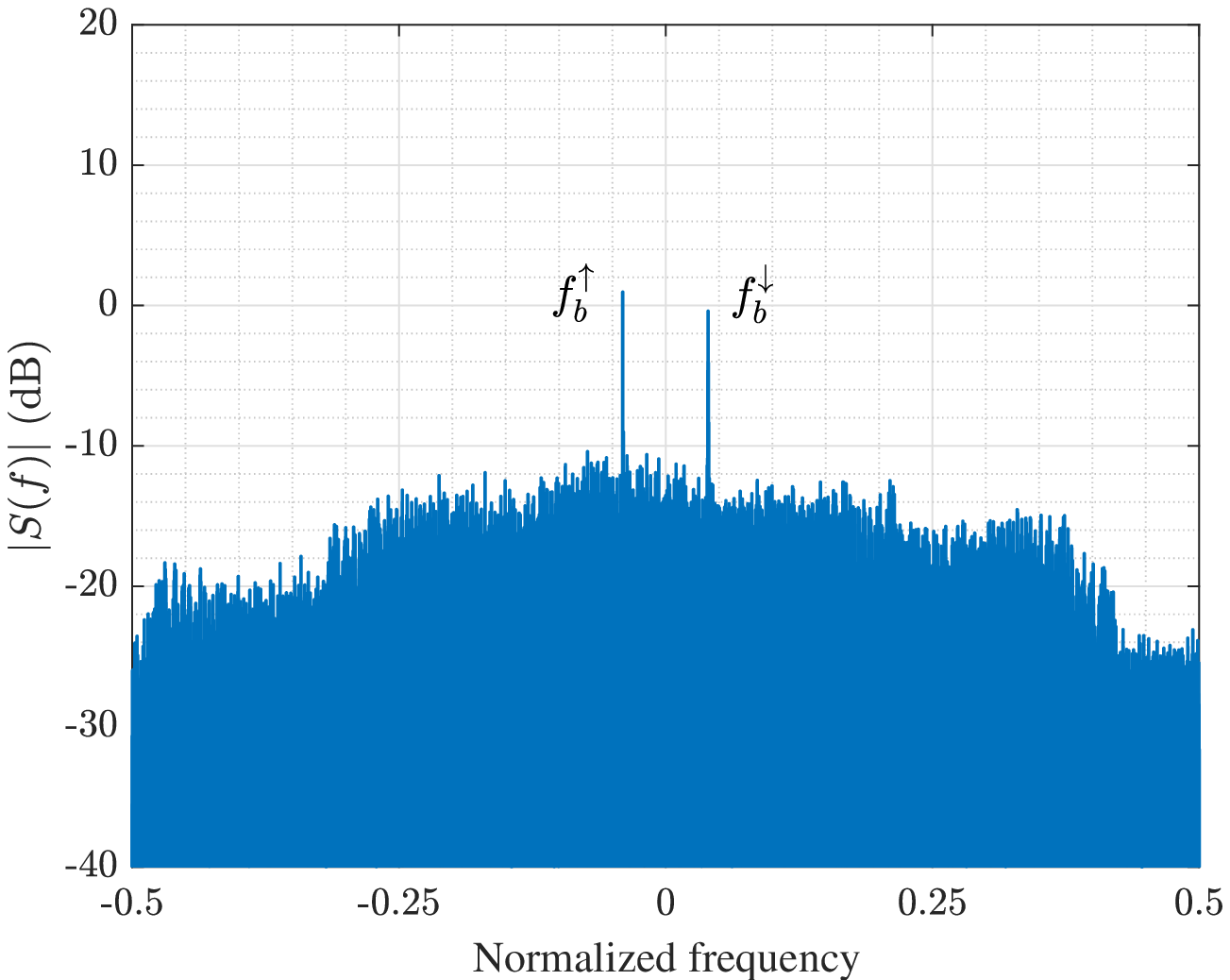}}
	\caption{\textcolor{black}{Results of three simulations in which a working radar is (a) not interfered; (b) interfered by another radar with the same chirp rate; (c) interfered by another radar with a different chirp rate. In the simulations, all signals are generated according to the signal models in Subsection II-B and -C. In (a), the received signal contains only the target echo and noise. The parameters of the working radar are: $B=200$ MHz, $f_m=76$ GHz, $T_c=50$ $\upmu$s, $T_f=1$ ms. In (b), the received signal contains the target echo, interference and noise. The interfering radar uses the same parameters as the working radar. In (c), the settings are the same as (b) except that the chirp interval of the interfering radar is $T'_c=20$ $\upmu$s.}}
	\label{fig:results}
\end{figure*}

\section{Radar Measurement}

\subsection{Range and Velocity}
\label{subsec:range_velocity}
\textcolor{black}{Range and velocity estimation of FMCW signal has been studied in \cite{Luo2013A}. In this subsection, we briefly explains how the range and velocity are estimated.}

\begin{figure}
	\centering
	\includegraphics[width=0.5\textwidth]{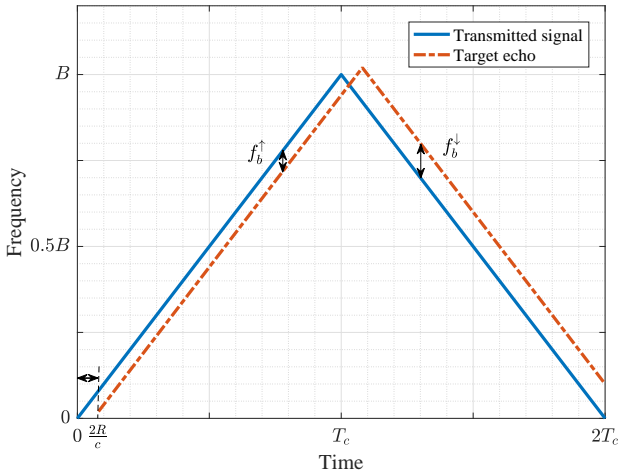}
	\caption{The instantaneous frequency of the transmitted signal and the target echo.}
	\label{fig:target_echo}
\end{figure}

\textcolor{black}{In FMCW signal processing, by mixing the received signal with the transimitted signal, we obtain the intermediate frequency (IF) signal:
\begin{equation}
r_{\text{IF}}(t) = r(t)\cdot\bar{s}(t),
\label{IF_signal}
\end{equation}
where $\bar{s}(t)$ indicates the conjugate of $s(t)$. The instantaneous frequency of $r_{\text{IF}}$ can expressed as
\begin{equation}
\frac{\partial\phi_{r_{\text{IF}}}(t)}{\partial t}=\frac{\partial \phi_r(t)}{\partial t}- \frac{\partial \phi_s(t)}{\partial t},
\end{equation}
where $\phi_{*}(t)$ and  $\frac{\partial\phi_{*}(t)}{\partial t}$ are the phase and instantaneous frequency of the signal, respectively.
As Fig. \ref{fig:target_echo} shows, the instantaneous frequency difference between the transmitted signal and the target echo, which is referred to as the beat frequency \cite{Brooker2007Mutual}, is approximately constant (the mathematical derivation can be found in APPENDIX \ref{app:A}).} The beat frequencies of the up-chirp and down-chirp can be expressed as \cite{Luo2013A}
\begin{equation}
\label{beat_freq1}
f_b^\uparrow \approx -\frac{B}{T_c}\cdot\frac{2D}{c}+\frac{2v}{c}f_m
\end{equation}
and
\begin{equation}
f_b^\downarrow \approx \frac{B}{T_c}\cdot\frac{2D}{c}+\frac{2v}{c}f_m,
\label{beat_freq2}
\end{equation}
respectively. \textcolor{black}{This implies that $r_{IF}(t)$ contains sinusoid components with the beat frequencies.} Hence, the estimates of $f_b^\uparrow$ and $f_b^\downarrow$ can be obtained by analyzing the spectrum
\begin{eqnarray}
R(f) = \mathcal{F}\left\{\bar{s}(t)r(t)\right\},
\label{received_spectrum}
\end{eqnarray}
where $\mathcal{F}$ represents the Fourier transform. \textcolor{black}{As Fig. \ref{fig:results}(a) shows, there appear peaks at the beat frequencies in the spectrum $R(f)$.} Then, the range and velocity can be calculated according to (\ref{beat_freq1})(\ref{beat_freq2}). 

\subsection{Interference}
\label{subsec:interference}
\textcolor{black}{If the working radar and the interfering radar use the same chirp rate, the interference signal (\ref{interference})(\ref{interference_delay}) has the similar form with the target echo (\ref{echo})(\ref{target_delay}) because $s'(t)=s(t)$. Hence, the interference also generates constant beat frequencies in a similar way with the target echo:
\begin{equation}
\label{beat_freq3}
f_b'^\uparrow \approx -\frac{B}{T_c}\cdot\frac{D'}{c}+\frac{v'}{c}f_m,
\end{equation}
and
\begin{equation}
f_b'^\downarrow \approx  \frac{B}{T_c}\cdot\frac{D'}{c}+\frac{v'}{c}f_m.
\label{beat_freq4}
\end{equation}
Therefore, additional peaks appear at the frequencies $f_b'^\uparrow$ and $f_b'^\downarrow$ in the spectrum $R(f)$,  resulting in ghost targets, which is shown in Fig. \ref{fig:results}(b).
To prevent ghost target, it is proposed in \cite{Luo2013A} that different radars use different chirp rates. In this way, the frequency difference between the interference and the transmitted signal is not constant but sweeps across the subband and the spectrum $H(f)$ is spread out within the bandwidth instead of causing additional peaks. Hence, it seems that the noise level is elevated, as shown in Fig. \ref{fig:results}(c).} The detailed mathematical derivation of the raise in the noise level can be found in APPENDIX \ref{app:B}.

From (\ref{received_signal})(\ref{IF_signal})(\ref{received_spectrum}), we have
\begin{eqnarray}
R(f)=E(f)+H(f)+N(f),
\end{eqnarray}
where
\begin{eqnarray*}
E(f) &=&\mathcal{F}\left\{e(t)\cdot\bar{s}(t)\right\},\\
H(f) &=& \mathcal{F}\left\{h(t)\cdot\bar{s}(t)\right\},\\
N(f) &=& \mathcal{F}\left\{n(t)\cdot\bar{s}(t)\right\}.
\end{eqnarray*}
Therefore,  when two radars uses different chirp rates, the interference power can be measured by the noise level, which is defined as:
\begin{eqnarray}
N_I = \int |H(f)+N(f)|^2 df.
\label{N_I_int}
\end{eqnarray}
The problem is to estimate $N_I$ using the spectrum of the received signal, $R(f)$, which contains target echoes as well as the received interference. Some robust estimation techniques have been developed to solve similar problems. For instance, in \cite{rohling1983radar}, ordered statistics is used to estimate the clutter power in existence of  heterogeneous samples. Likewise, we use ordered statistics to estimate the noise level. 

\textcolor{black}{In signal processing, the spectrum $R(f)$ in (\ref{received_spectrum}) is usually obtained by the fast Fourier transform (FFT). This equals sampling $R(f)$ with a sampling interval $\Delta f$, which is determined by the sampling rate and the number of FFT points. Then, we obtain a sequence $\{R_m\} (0\le m\le M_f-1)$, where $M_f$ is the number of FFT points and
\begin{equation}
R_m=R(m\Delta f).
\label{R_m}
\end{equation}}
Then, by sorting the sequence according to decreasing amplitude, we obtain a new sequence $\{\hat{R}_m\}$:
\begin{equation}
|\hat{R}_0| \ge |\hat{R}_1| \ge...\ge |\hat{R}_{M_f-1}|,
\label{sort_R_m}
\end{equation}
where $\hat{R}_m$ is the element in $\{R_m\}$ with the $m$th largest amplitude. \textcolor{black}{As explained in Subsection \ref{subsec:range_velocity}, the power of the target echo mainly concentrate on the peaks of the spectrum $R(f)$. Hence, by discarding the greatest  $K$ samples, we can obtain an estimate of the noise level:
\begin{equation}
\hat{N}_{I} = \frac{M_f}{M_f-K}\sum_{m=K}^{M_f-1} |\hat{R}_m|^2\Delta f,
\label{N_I}
\end{equation}
where $K$ is the number of discarded values. The value of $K$ can be approximately selected as
\begin{equation}
K\approx n_{\max}\cdot\frac{B}{T_c}\cdot\frac{2l_{\max}}{c}\cdot\frac{1}{\Delta f},
\end{equation}
where $l_{\max}$ is the maximum target size, $n_{\max}$ is the maximum number of targets and  $\dfrac{B}{T_c}\cdot\dfrac{2l_{\max}}{c}$ indicates the frequency range that a target with size $l_{\max}$ occupies.
In practice, the value of $K$ can be set larger than the equation above to ensure that all peaks corresponding to the target echo are discarded. The discarded values except for the target echo have little influence on the estimation of $N_I$ because the estimation in (\ref{N_I}) is an average of all values and moreover, $M_f\gg K$ .}

\begin{figure}[!t]
 	\centering
 	\includegraphics[width=0.5\textwidth]{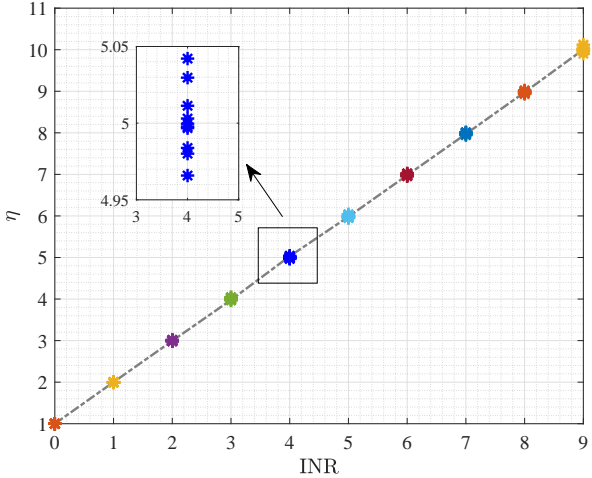}
 	\caption{Relative noise level versus INR. At each value of INR, we simulate 10 cases where two radars randomly select two different chirp rates. The inner subfigure is a magnified version of the plot within the box. }
 	\label{fig:noise_level_INR}
 \end{figure}

Denote the interference-free noise level as
\begin{equation}
N_{IF}=\int |N(f)|^2 df, 
\label{N_F_int}
\end{equation}
which is related to the receiver noise power and can be regarded as known. Let
\begin{equation}
\eta = \frac{\hat{N}_I}{N_{IF}}
\label{eta}
\end{equation}
denote the estimation of the relative noise level.
In Fig. \ref{fig:noise_level_INR}, we plot the relative noise level versus the interference-to-noise ratio (INR). The INR is defined as 
\begin{equation}
\textrm{INR} = \frac{P_I}{\sigma^2},
\end{equation}
where $\sigma^2$ is the receiver noise power. At each value of INR, we simulate 10 cases where two radars randomly select two different chirp rates. \textcolor{black}{In each simulation, the received signal power to the noise power ratio is set as $4$ and the interference power varies. In the estimation of $\eta$, we set $K=20$.} Fig. \ref{fig:noise_level_INR} conveys two messages. First, $\eta$ increases approximately linearly with INR. \textcolor{black}{More specifically, we have $\eta\approx \textrm{INR}+1$, because $N_I$ is proportional to the interference plus the noise power and $N_{IF}$ to the noise power.} Second, $\eta$ is almost irrelevant to different combinations of chirp rates. Therefore, $\eta$ is a suitable measurement for the interference power in our problem.
 
\textcolor{black}{Now, by exploiting the spectrum of the received signal $R(f)$ according to (\ref{R_m}-\ref{eta}), we obtain an estimation of the relative noise level, $\eta$, as the radar measurement for the interference power. In this way, radar only monitors the subband on which it transmits.}

\section{Spectrum allocation using RL}



\textcolor{black}{At the beginning of this section, we clarify some notations. The subscript $t$ represents discrete time step and the superscript $i$ represents the car index and $i\in\{1,2,...,N\}$.}

\begin{figure*}[!t]		
	\centering 
	\includegraphics[width=\textwidth]{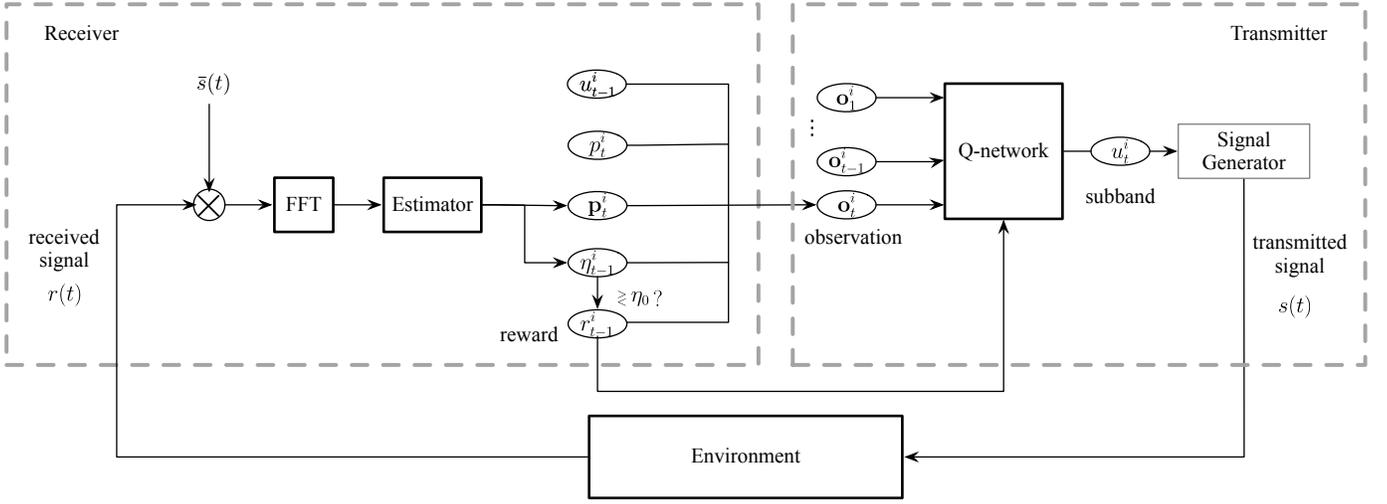}	
	\caption{Outline of the RL-based spectrum allocation approach for automotive radar.}    
	\label{fig:RL_CR}
\end{figure*}

The outline of the RL-based spectrum allocation approach for automotive radar is shown in Fig. \ref{fig:RL_CR}. First, by processing the received signal from last time step, the receiver constructs the current observation $\mathbf{o}^i_t$. The construction of  $\mathbf{o}^i_t$ will be described in detail in Subsection \ref{subsec:observation}. Then, the transmitter employs a Q-network to choose a subband $u_t^i$ by aggregating the historical observations $\{\mathbf{o}^i_t,\mathbf{o}^i_{t-1},...,\mathbf{o}^i_1\}$. In the meantime, the receiver also gives feedback to the transmitter in the form of a reward signal $r^i_{t-1}$, which is evaluated based on the relative noise level $\eta^i_{t-1}$. The reward acts as a tutor guiding the Q-network to adjust parameters to generate a better subband selecting policy.


\subsection{Reward and Receiver Observation}
\label{subsec:observation}
The reward is defined in terms of the relative noise level $\eta_t^i$. If $\eta_t^i$ is below a predefined threshold $\eta_0$, the transmission is regarded as successful and the corresponding reward is $1$; otherwise, the reward is $0$, i.e.
\begin{equation}
r_t^i = \left\{\begin{array}{ll}
1&\eta_t^i<\eta_0\\
0&\eta_t^i\ge\eta_0
\end{array}\right..
\end{equation}

The observation which Car $i$ acquires at time step $t$ is 
\begin{equation}
\mathbf{o}^i_t=\left[u^i_{t-1},\  r^i_{t-1},  \eta^i_{t-1},\  p^i_t, \ \mathbf{p}^i_t\right],
\label{observation1}
\end{equation} 
where
\begin{itemize}[ ]
	\item $u^i_{t-1}$: the last subband on which Car $i$ transmitted;\vspace{2pt}
	\item $r^i_{t-1}$: the last reward which Car $i$ received;\vspace{2pt}
	\item $\eta^i_{t-1}$: the relative noise level in last time step;\vspace{2pt}
	\item $p^i_t$: \quad the position of Car $i$;\vspace{2pt}
	\item $\mathbf{p}^i_t$: \quad the estimated positions of cars in front of Car $i$.
\end{itemize}  
The car's own position $p_t^i$ can be acquired from its onboard sensors such as the global positioning system (GPS) and inertial navigation system (INS). The estimated position $\mathbf{p}^i_t$ is a two dimensional vector:
\begin{equation}
	\mathbf{p}^i_t = [p^i_{S,t}, p^i_{D,t}],
\end{equation}
where $p^i_{S,t}, p^i_{D,t}$ are the position of the nearest car in front of Car $i$ in the same and different lane, respectively. The vector $\mathbf{p}^i_t$ is calculated according to the radar measurement of the range and velocity, which is accompanied with estimation errors caused by noise and interference.

\subsection{Q-network in Transmitter}
\label{subsec:qnetwork}
In this subsection, first, we recap the RL and Q-learning algorithm. Then, we show how the Q-network is specified for the spectrum allocation problem.

In RL, an agent learns how to choose actions by receiving rewards from an unknown environment \cite{Sutton1998Reinforcement}. Let $\mathbf{s}_t$, $a_t$ and $r_t$ denote the state of the environment, the action of the agent and the reward it receives at time step $t$. At each time step, action $a_t$ is determined by environment state $\mathbf{s}_t$ following a policy $\pi$, which is a mapping from the state space to the action space, by trying to maximize a discounted sum of future rewards:
\begin{equation}
	G_t = r_t+\gamma r_{t+1}+\gamma^2 r_{t+2}+... ,
\end{equation}
where $\gamma\in[0,1]$ is the discounting factor. The factor $\gamma$ reflects how much we consider the influence of the current action on the future. An extreme example is $\gamma=0$, which corresponds to the case where the agent aims to maximize the immediate reward $r_t$.
The Q-function is defined as the expectation of $G_t$ after taking action $a_t$ under environment state $\mathbf{s}_t$ following policy $\pi$:
\begin{equation}
Q_{\pi}(\mathbf{s}_t,a_t)=\mathbb{E}\left\{G_t|\mathbf{s}_t, a_t,\pi\right\},
\end{equation} 
where the expectation is taken over the probabilistic sequence, $\mathbf{s}_{t+1}, a_{t+1}, r_{t+1}, \mathbf{s}_{t+2}, a_{t+2}, r_{t+2}, ... $, following policy $\pi$. Learning the optimal policy equals finding the optimal Q-function:
\begin{eqnarray}
Q^*(\mathbf{s}_t,a_t) = \max_{\pi}Q_{\pi}(\mathbf{s}_t,a_t).
\label{Q_optimal}
\end{eqnarray}
Then, the best action can be determined by the optimal Q-function:
\begin{equation}
a^*_t = \arg\max\limits_{a'}Q^*(\mathbf{s}_t,a').
\end{equation}

The Q-learning algorithm provides an iterative way to estimate the optimal Q-function even when an explicit model of the environment is unavailable. Each iteration is based on an experience of the agent, which is represented by a quadruple, $\left(\mathbf{s}_t,a_t,r_t,\mathbf{s}_{t+1}\right)$. The iteration is performed as \cite{Sutton1998Reinforcement}:

\begin{eqnarray}
\nonumber
Q(\mathbf{s}_t, a_t) &\leftarrow&Q(\mathbf{s}_t,a_t)\\\nonumber
&&+\alpha_t\left[r_t+\gamma\max\limits_{a'}Q(\mathbf{s}_{t+1},a')-Q(\mathbf{s}_t,a_t)\right],\\
\label{Q_iteration}
\end{eqnarray}
where $\alpha_t$ is the learning step size. As the iteration in (\ref{Q_iteration}) is limited to cases where the state and action space are low dimensional and discrete, a neural network is usually used to approximate the Q-function \cite{Riedmiller2005Neural}. \textcolor{black}{The Q-function approximated by a neural network is referred to as Q-network hereafter.}

In our problem, as $\mathbf{s}_t$ is not fully observable, using a single observation to represent the environment state is inadequate. To construct a more complete environment state, the historical observations are aggregated: 
\begin{equation}
\mathbf{s}^i_t = \left[\mathbf{o}^i_1, \mathbf{o}^i_2,..., \mathbf{o}^i_t\right],
\end{equation}
where $\mathbf{s}^i_t$ denotes the constructed environment state by Car $i$. Here, we use an LSTM recurrent neural network to approximate the Q-function since the LSTM structure is capable of memorizing the past by maintaining a hidden state \cite{Hochreiter1997Long}. The Q-network for Car $i$ is denoted as  $Q^i(\mathbf{o}^i_t, \mathbf{h}^i_{t-1}, u^i_t; \mathbf{w}^i)$, where $\mathbf{w}^i$ is the network parameter, $u^i_t$ is the chosen subband by Car $i$ at time step $t$ and $\mathbf{h}^i_{t-1}$ is the hidden state extracted from past observations $\mathbf{o}^i_1,\mathbf{o}^i_2,...,\mathbf{o}^i_{t-1}$. Using the LSTM network, the constructed environment state by Car $i$ can be equally represented by:
\begin{equation}
\mathbf{s}^i_t = \left[\mathbf{o}^i_t,\mathbf{h}^i_{t-1}\right].
\end{equation}



\begin{figure*}
\begin{minipage}{0.49\textwidth}
	\centering
	\includegraphics[width=\textwidth]{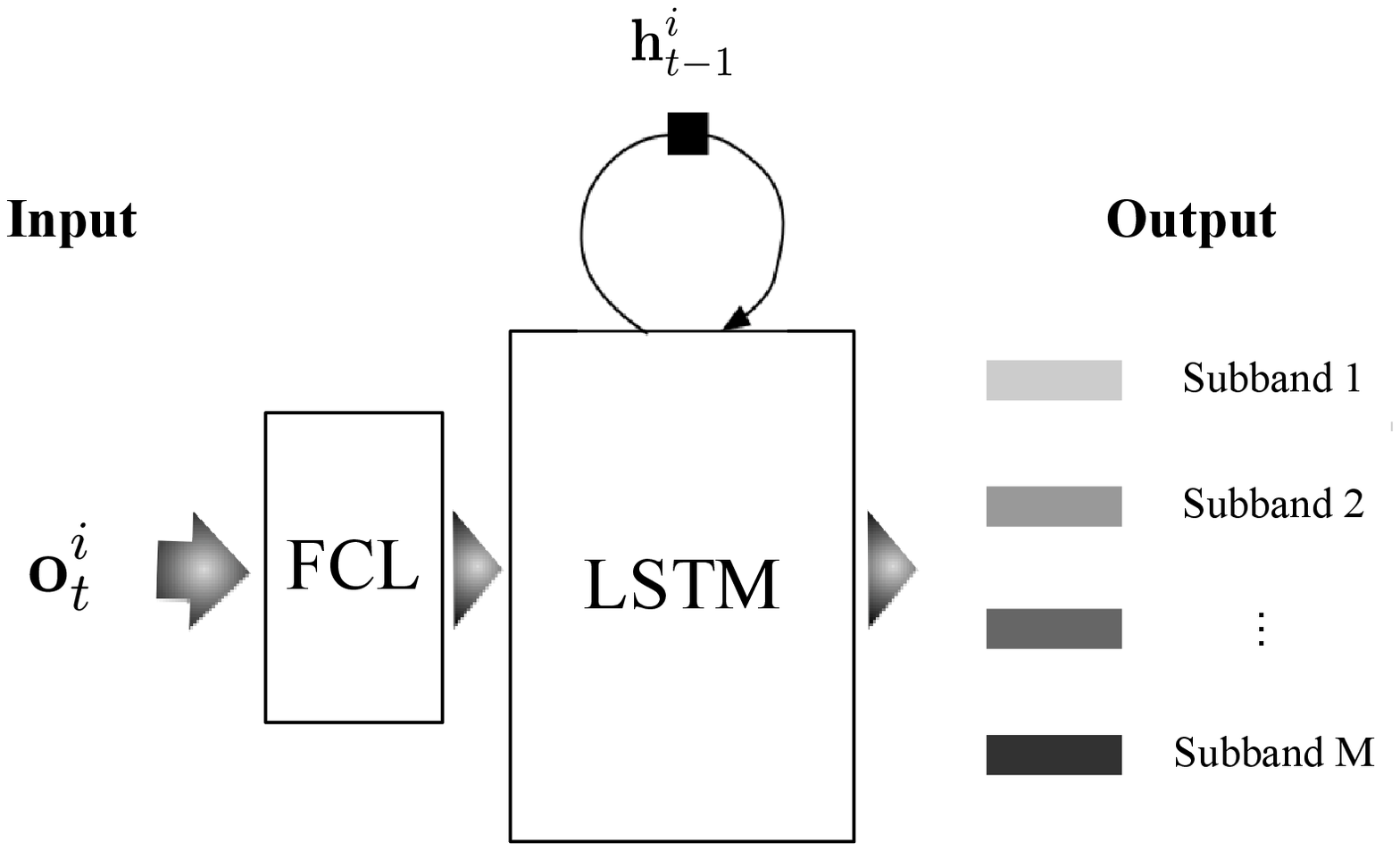}
	\caption{The Q-network architecture. The black square stands for a time-step delay.}
	\label{fig:q_network_structure}
\end{minipage}
\hfill
\begin{minipage}{0.49\textwidth}
	\centering
	\includegraphics[width=\textwidth]{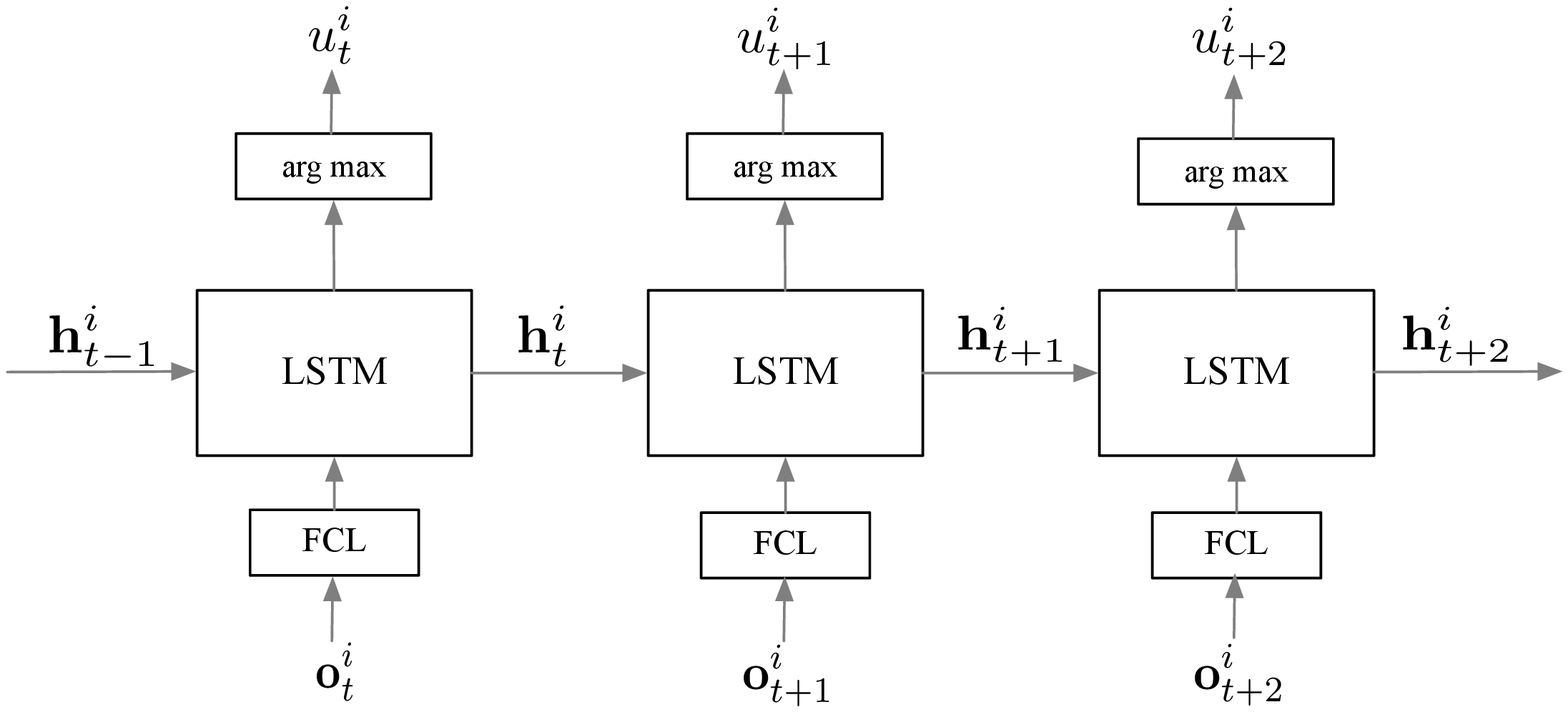}
	\caption{Unfolded representation of the Q-network, showing a subband is determined by combining the present and past observations.}
	\label{fig:q_network_unfolding}
\end{minipage}
\end{figure*}





The loss function of the $i$-th Q-network is defined as 
\begin{eqnarray}
\label{loss_function}
\mathcal{L}^i(\mathbf{w}^i)=\mathbb{E}\left\{\left( y_t^{i}-Q^i\left(\mathbf{o}^i_{t},\mathbf{h}^i_{t-1},u^i_t;\mathbf{w}^i\right)\right)^2\right\},
\end{eqnarray}
where the expectation is taken over each experience $(\mathbf{o}^i_{t},\mathbf{h}^i_{t-1},u^i_t,r^i_t,\mathbf{o}^i_{t+1},\mathbf{h}^i_{t})$, and $y_t^{i}$ is the target value \cite{Sutton1998Reinforcement}:
\begin{equation}
y_t^{i}=r^i_t+\gamma\max_{u'}Q^i\left(\mathbf{o}^i_{t+1},\mathbf{h}^i_{t},u';\mathbf{w}^{i-}\right),
\label{target_value2}
\end{equation}
where $\mathbf{w}^{i-}$ represents the network parameters before the update. The loss function is to minimize the error between the Q-function and target values over experiences of all radars. Then, the gradient descent step is performed to finish the update:

\begin{equation}
\mathbf{w}^i = \mathbf{w}^{i-}-\beta \frac{\partial \mathcal{L}^i(\mathbf{w}) }{\partial\mathbf{w}}\bigg|_{\mathbf{w} = \mathbf{w}^{i-}},
\label{gradient_descent}
\end{equation}
where $\beta$ is the learning rate.

\subsection{Network Architecture}
\label{subsec:q_network}

The Q-network architecture in our problem is shown in Fig. \ref{fig:q_network_structure}. A fully connected layer (FCL) transforms $\mathbf{o}_t^i$ into inputs of the LSTM layers.  The LSTM layers maintain a hidden state $\mathbf{h}_{t-1}^i$, which is extracted from past observations before time step $t$.
Each element of the output represents the Q-function value of the corresponding subband. Fig. \ref{fig:q_network_unfolding} shows the unfolded representation of the Q-network. At time $t$, subband $u^i_t$ is chosen by combining the current observation $\mathbf{o}^i_t$ and hidden state $\mathbf{h}^i_{t-1}$. Then $\mathbf{h}^i_{t-1}$ evolves into $\mathbf{h}^i_{t}$ by incorporating the new observation $\mathbf{o}^i_t$. In this way, the subband at each time step is determined by both the present and past observations.

\subsection{Network Training}
\label{subsection:train}

In this subsection, some details of the network training are given.

\textcolor{black}{In the training, a time step is equal to a transmission period $T$. An episode consists of a number of successive time steps randomly taken from the dynamic scenario in which cars are traveling in two lanes.} In \cite{Volodymyr2015Human}, several techniques are employed to train the deep Q-network with stability. In our problem, the following listed techniques are adopted to train the Q-networks.

\begin{itemize}
\item \emph{Experience replay} with \emph{batch learning}. During the training, each experience,  $e^i_t = \left(\mathbf{o}^i_{t},\mathbf{h}^i_{t-1}, u^i_t,r^i_t,\mathbf{o}^i_{t+1},\mathbf{h}^i_t\right)$, is stored in memory. The memory only stores experiences of the recent $200$ episodes. In each update, a batch of experiences are drawn from the memory. A batch is formed as (\ref{batch2}) shows.
\begin{equation}
\left\{\begin{array}{cccc}
e_{t_1}^{i}&e_{t_1+1}^{i}&\cdots&e_{t_1+P-1}^{i}\\
e_{t_2}^{i}&e_{t_2+1}^{i}&\cdots&e_{t_2+P-1}^{i}\\
\vdots&\vdots&\ddots&\vdots\\
e_{t_K}^{i}&e_{t_K+1}^{i}&\cdots&e_{t_K+P-1}^{i}\\
\end{array}\right\}
\label{batch2}
\end{equation}
Each row in  (\ref{batch2}) is obtained following the two procedures: first, randomly picking an episode from the memory; then, randomly choosing a sequence of $P$ successive experiences from the whole episode. The randomization among different rows improves training stability because it breaks the correlation of experience sequences \cite{Volodymyr2015Human}. \textcolor{black}{The batch is used for the gradient descent step (\ref{gradient_descent}) in which the gradient ${\partial\mathcal{L}(\mathbf{w})}/{\partial\mathbf{w}}$ is calculated on the batch. The gradient calculation can be easily realized with the Tensorflow framework.}

\item $\epsilon$-\emph{greedy policy}. At each time step, each car chooses a subband corresponding to the greatest Q-function value or otherwise a random one with small exploring probability $\epsilon$, i.e.
\begin{equation}
u^i_t=\left\{\begin{array}{ll}
\arg\max\limits_{u'} Q^i(\mathbf{o}^i_t,\mathbf{h}^i_{t-1},u';\mathbf{w}^i)& a \ge \epsilon\\
\textrm{a random subband} &a<\epsilon
\end{array}\right.,
\label{epsilon_greedy}
\end{equation}
where $a$ is one realization of a uniformly distributed random variable ranged from 0 to 1.

\item \emph{Double networks}. \textcolor{black}{To enhance training stability, a separate network, called target network and denoted as $\hat{Q}^i$, is introduced in the training procedures. In (\ref{target_value2}), the target value is generated by the Q-network itself, which may cause training instability.  
Instead, the target network is used to produce the target value in the training and (\ref{target_value2}) is replaced by
\begin{equation}
y_t^{i}=r^i_t+\gamma\max_{u'}\hat{Q}^i\left(\mathbf{o}^i_{t+1},\mathbf{h}^i_{t},u';\mathbf{\hat{w}}^{i}\right),
\label{target_value3}
\end{equation}
where $\mathbf{\hat{w}}^{i}$ is the network parameter of $\hat{Q}^i$. The target network $\hat{Q}^i$ has the same structure with $Q^i$. Unlike the Q-network which is updated every time step, the target network is updated every few time steps. In an update, the Q-network parameter $\mathbf{w}^i$ is assigned to the target network parameter $\hat{\mathbf{w}}^i$. Between two updates, the target network parameter is held fixed. The number of steps between two updates, i.e. the training cycle of the target network, is denoted as $C$.}
\end{itemize}

The Q-network training algorithm is summarized in Algorithm 1.

\begin{algorithm}[!htbp]
	\SetAlgoNoLine
	\caption{Q-Network Training Algorithm }
	Set up two set of networks, $\{Q^i\}$ and $\{\hat{Q}^i\}$.\\
	Initialize $Q^i$ and $\hat{Q}^i$ with the same random parameter $\mathbf{w}^-$.\\
	\For{$\textrm{episode} = 1:N_e$}{
		Initialize hidden state $\mathbf{h}^i_0$ and observation $\mathbf{o}_1^i$, $\forall i\in\{1,2,...,N\}$.\\
		\For{$t=1:T$}{
			\For{Car $i=1:N$}{
				Feed observation $\mathbf{o}^i_t$ to network $\mathbf{Q}^i$ to get a set of action values $\{Q^i(\mathbf{o}^i_t,\mathbf{h}^i_{t-1},u;\mathbf{w})\}(u=1,2,...,M)$ and the hidden state $\mathbf{h}_t^i$.\\
				Choose a subband $u^i_t$ as (\ref{epsilon_greedy}).\\
				Get a reward $r^i_t$.\\
				Obtain new observation $\mathbf{o}^i_{t+1}$\\
				Store $e^i_t = \left(\mathbf{o}^i_{t},\mathbf{h}^i_{t-1}, u^i_t,r^i_t,\mathbf{o}^i_{t+1},\mathbf{h}^i_t\right)$  into memory.\\
				Form a batch as (\ref{batch2}).\\
				Calculate target values $y^i_t$ using network $\hat{Q}^i$.\\
				$\mathbf{w}^i \leftarrow \mathbf{w}^i-\beta \frac{\partial \mathcal{L}^i(\mathbf{w}) }{\partial\mathbf{w}}$.\\
				Every $C$ time steps, $\hat{Q}^i=Q^i$.
				}
			}
		}
\label{LSTM_training_algorithm}
\end{algorithm}

\section{Simulations}
\label{sec:simulation}

In this section, simulation results are presented to verify the proposed approach. First, the simulation setup is described. Then, two contrasting approaches are introduced. Last, simulation results are provided along with the corresponding discussions .

\textcolor{black}{The simulations are implemented with Python. The networks used in the simulations are built and trained with the Tensorflow framework.}

\subsection{Simulation Setup}
\label{subsec:setup}
\textcolor{black}{Training Q-networks can be time-consuming. In our case, a training episode consists of around 20 to 200 time steps and it takes thousands of episodes for the Q-networks to converge. Hence, in many RL related applications, the Q-networks are first trained offline in a synthesized environment and then tested in other environments to see if the trained Q-networks are capable of generalization. This practice is also adopted in our simulations. More specifically, we first train the Q-networks in a simulated environment with a relatively simple scenario model. Then, we test the trained Q-networks with a more complex scenario model.}

The simulated scenario for training is constructed as Fig. \ref{fig:scenario} shows. The flow of traffic is modeled by a truncated exponential distribution \cite{Liu2011Throughput}. Under this model, the distance $l$ between any two adjacent cars satisfies the following distribution: 
\begin{equation}
\label{truncated_exponential_distribution}
p(l)=\left\{\begin{array}{ll}
\lambda\cdot\frac{1}{\rho}\exp(-\frac{l}{\rho})&d_{\min}\le l\le d_{\max}\\
0&\textrm{otherwise}\end{array}\right.,
\end{equation}
where $\rho$ is the intensity parameter, and $\lambda$ is a normalizing coefficient to assure that the integral of $p(l)$ equals 1. Cars in each lane are assumed to travel at constant velocity, i.e., $v_1$ and $v_2$, respectively. The detailed scenario settings are shown in TABLE \ref{tab:scenario_settings}. In the simulations, if $\textrm{INR}\le 10$ (10 dB), it is taken as a successful transmission. As shown in Subsection III-B, $\eta\approx \textrm{INR}+1$. Hence, we set the threshold $\eta_0$ as $11$.

\textcolor{black}{In the testing scenario, the motion of each car is modeled by a probabilistic cellular automaton model \cite{SchreckenbergDiscrete}, which is a widely used model in traffic flow simulations. Let $v_{\max,i}$ ($i=1,2$ indicate the two lanes, respectively) denote the maximum velocity for the two lanes. At each time interval $t_v$, the velocity of an arbitrary car is updated by:
\begin{itemize}
	\item [1)] \textbf{Acceleration}: If the velocity $v$ of a car is lower than $v_{\max,i}$, then the car speeds up by $v=v+\Delta v$, where $\Delta v$ is a predefined velocity increment.
	\item [2)] \textbf{Slowing down}: If the distance $l$ to the next car ahead in the same lane is not larger than $d_{\min}$, then the car slows down by  $v=v-\Delta v$.
	\item [3)] \textbf{Randomization}: With probability $p_{\textrm{sd}}$, the car slows down by  $v=v-\Delta v$ if  $v\le\Delta v$.
\end{itemize} }

\begin{table}[!htbp]
	\caption{Scenario settings}
	\centering
	\begin{tabular}{p{0.07\textwidth}p{0.28\textwidth}p{0.07\textwidth}}
		\textbf{Notation}&\textbf{Description}&\textbf{Value}\\
		\toprule[1pt]
		$T$&Transmission period/a time step (ms)&$100$\\
		$T_f$&Frame duration (ms)&$5$\\
		$T_c$&Chirp interval ($\upmu$s)& $10\sim100$\\		
		$P_L$&Transmitting power of long-range radar (dBmW)&$25$\\
		$P_S$&Transmitting power of short-range radar (dBmW)&$15$\\
		$A_e$&Effective area (mm$^2$)&$5$\\
		$G_t$&Antenna gain&$48$ dB\\
		$g$&Decaying coefficient&$0.1$\\
		$v_{1,\max}/v_1$&(Maximum) velocity of cars on Lane 1 (m/s)&  $30$ \\
		$v_{2,\max}/v_2$&(Maximum) velocity of cars on Lane 2 (m/s)&  $-25$ \\
		$\Delta v$&Velocity increment (m/s)&$5$\\
		$t_v$&Velocity updating interval (s)&$0.5$\\
		$\eta_0$&Relative noise level threshold&11\\
		\toprule[1pt]
	\end{tabular}
	\label{tab:scenario_settings}
\end{table}


The hyper parameters used in training the Q-networks are shown in TABLE \ref{tab:network_parameters}.
\begin{table}[!htbp]
	\centering
	\caption{Parameters in training the Q-networks}
	\begin{tabular}{llc}
		\textbf{Description}&\multicolumn{2}{c}{\textbf{Value}}\\
		\toprule[1pt]
		& 	\textbf{Layer} & 	\textbf{Number of neurons}\\
		& &\\
		& Input &$7$\\
		& FCL &$30$\\
		Network & LSTM1 &$30$\\
		architecture& LSTM2 &$30$\\
		& LSTM3& $20$\\
		& LSTM4 &$10$\\
		& Output &$M$\\
		 \hline
		 & &\\
		 Time step&\multicolumn{2}{c}{1 ms}\\
		Episode length &\multicolumn{2}{c}{around $20\sim200$ time steps}\\
		Batch size $K\times P$ &\multicolumn{2}{c}{$40\times 20$}\\
		Exploring probability $\epsilon$&\multicolumn{2}{c}{$0.05$}\\
		Target network training cycle $C$&\multicolumn{2}{c}{$20$}\\
		\toprule[1pt]
	\end{tabular}
	\label{tab:network_parameters}
\end{table}


\subsection{Contrasting Approaches}
The first contrasting approach is the random policy, which is to randomly select a subband with equal probability at each time step. The second is a commonly used method in DSA, the myopic policy \cite{Zhao2008On}. The myopic policy aims to select a subband which is most likely to be unused in the next time step. However, it requires prior knowledge of the transition probability of each subband. In \cite{Zhao2008On}, a practical realization of the myopic policy is given without knowing the transition probability. To make it more efficient in our multi-radar problem, we make some modifications. In the original myopic policy, all the subbands are kept in a predefined priority order. The user keeps using one subband if the result is a success. Otherwise, it switches to the next subband according to the priority order. In the modified myopic policy, the order is not needed. Radar switches to a random subband when failure occurs. The modified myopic policy outperforms the original version in our problem. Under the modified myopic policy, the subband chosen at time step $t$ is 
\begin{equation}
u^i_t=\left\{\begin{array}{ll}
u^i_{t-1}&\eta^i_t<\eta_0\\
\textrm{a random subband}&\textrm{otherwise}
\end{array}\right..
\end{equation}

\subsection{Performance Metric}
\label{subsec:metric}
\textcolor{black}{
To evaluate the proposed and contrasting approaches, we use success rate as a metric. Success rate is the percentage of the successful transmissions, expressed as
\begin{equation}
\xi=\frac{1}{N}\sum_{i=1}^{N}\frac{1}{N_t}\sum_{t=0}^{N_t}\mathbb{I}_{\eta_0}(\eta_i^t),
\label{xi}
\end{equation}
where $N_t$ is the total number of transmissions and $\mathbb{I}_{\eta_0}(\eta_i^t)$ is defined as
\begin{equation}
\mathbb{I}_{\eta_0}(\eta_i^t)=\left\{\begin{array}{ll}
	1&0\le \eta_i^t \le \eta_0 \\
	0&\textrm{elsewhere}\end{array}\right..
	\label{I}
\end{equation}}

\subsection{Results and Discussions}
\label{subsec:results}
\begin{figure*}[t]
	\centering
	\subfloat[$N=6,M=2$]{\includegraphics[width=0.5\textwidth]{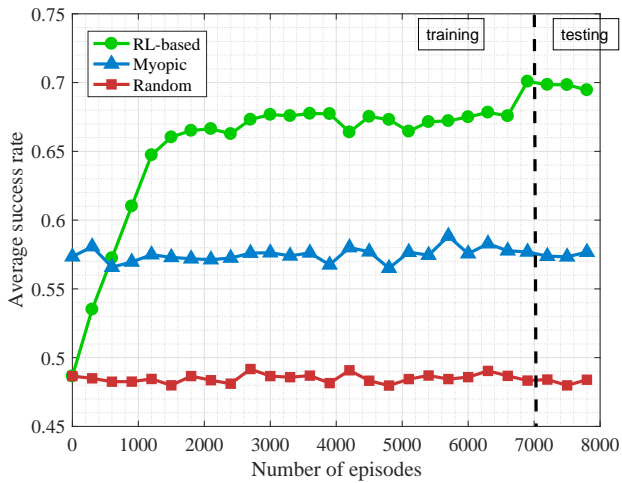}}
	\subfloat[$N=8,M=3$]{\includegraphics[width=0.5\textwidth]{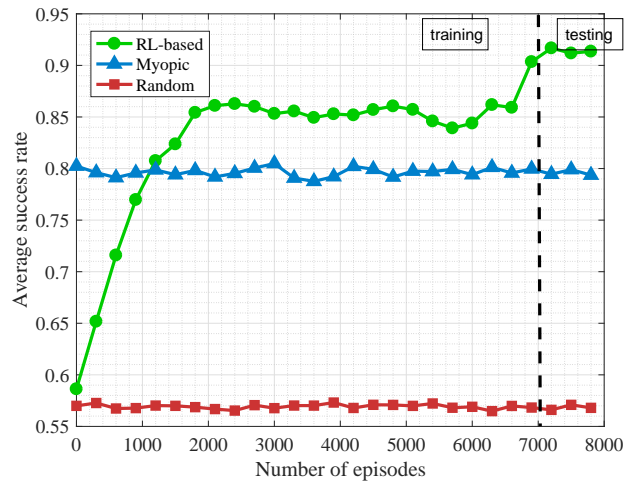}}
	\caption{Success rate of the three approaches versus the number of episodes in two scenarios. In the first $7,000$ episodes, the Q-networks are trained with the uniform motion model. In the last $1000$ episodes, the trained networks are tested with the probabilistic cellular automaton model.}
	\label{fig:learning_curves}
\end{figure*}

In Fig. \ref{fig:learning_curves}, a comparison of the success rate achieved by the three approaches is drawn in two scenarios where $N=6,M=2$ and $N=8,M=3$ ($N, M$ are the number of cars and subbands), respectively. In the first $7,000$ episodes, the Q-networks are trained with the uniform motion model. In the last $1,000$ episodes, the trained networks are tested with the probabilistic cellular automaton model.
Apparently, the myopic policy performs better than the random policy. In Fig. \ref{fig:learning_curves} (a), the average success rate achieved by the myopic and random policy are $58\%$ and $47\%$; in Fig. \ref{fig:learning_curves} (b), they are $80\%$ and $57\%$. As the RL-based curves show, the success rate gradually increases and then becomes stable during the training. In the beginning, the RL-based approach is close to the random policy and outperformed by the myopic policy. As the learning proceeds, the RL-based approach gradually surpasses the myopic policy. In the testing, the trained Q-networks achieve success rate of 70\% in Fig. \ref{fig:learning_curves} (a) and $90\%$ in Fig. \ref{fig:learning_curves} (b), both realizing around $10\%$ success rate improvement over the myopic policy in both scenarios. 
It should be noted that during the training each radar maintains an exploring probability $\epsilon=0.05$ to randomly choose a subband. In the testing, each radar chooses subbands according to the learned Q-networks without exploring, i.e. $\epsilon=0$. This accounts for a small rise in the success rate in the last $1000$ episodes.  

\begin{figure*}[t]
	\centering
	\includegraphics[width=1\textwidth]{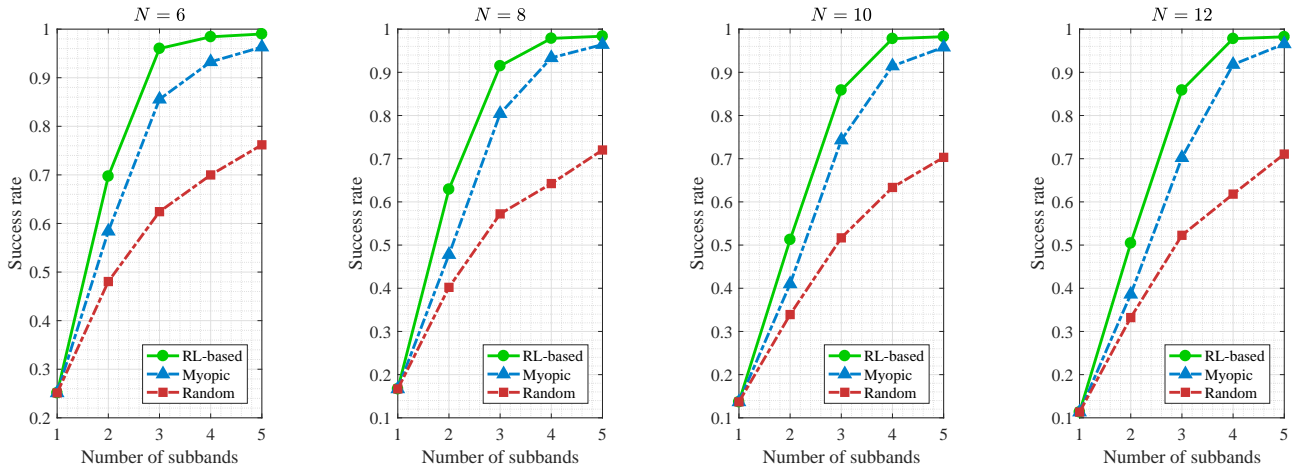}
	\caption{Success rate achieved by the three approaches versus the number of subbands when the number of cars is fixed. }
	\label{fig:SR}
\end{figure*}

\begin{figure*}[h]
	\centering
	\subfloat[RL-based v. random]{\includegraphics[width=0.5\textwidth]{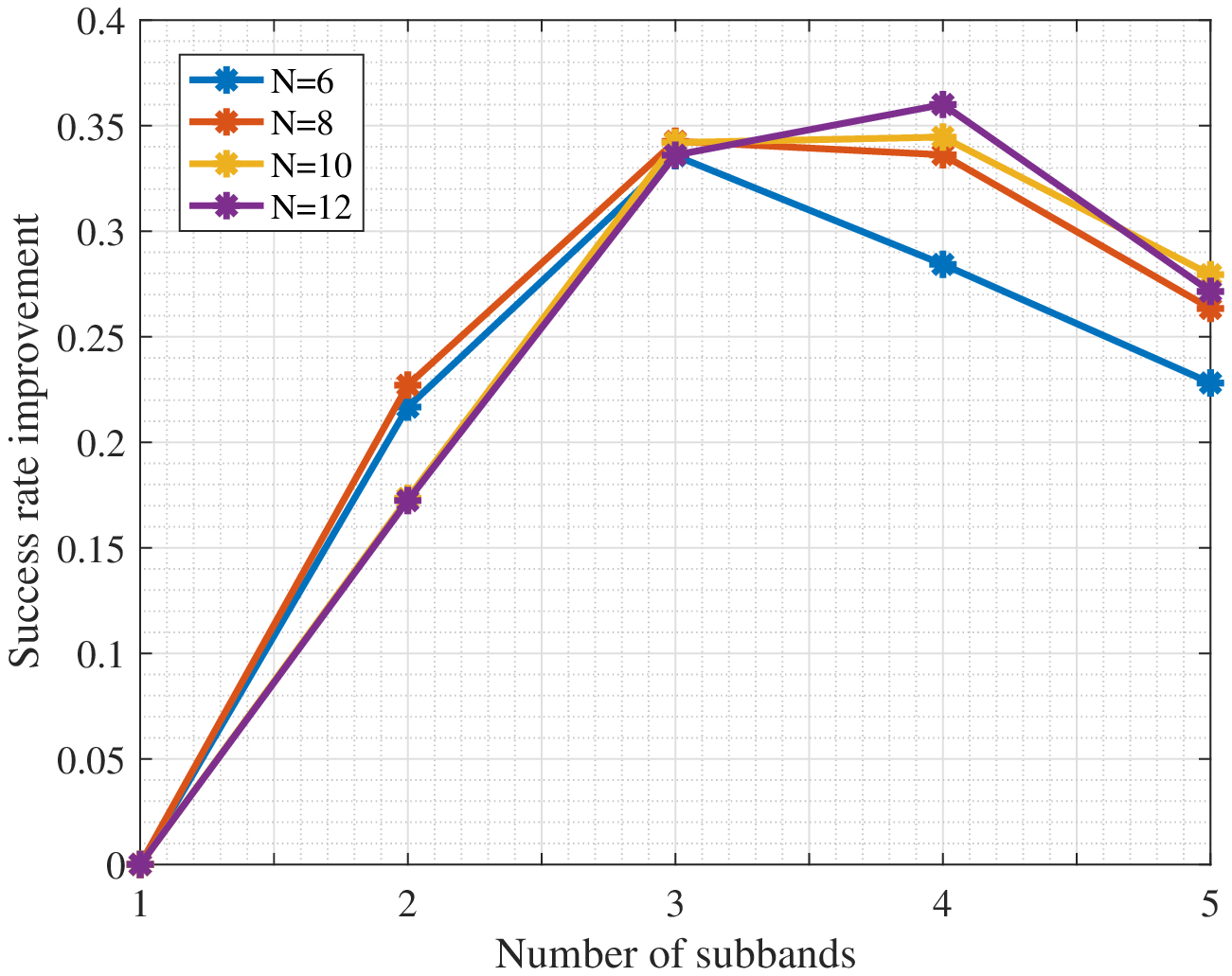}}
	\subfloat[RL-based v. myopic]{\includegraphics[width=0.5\textwidth]{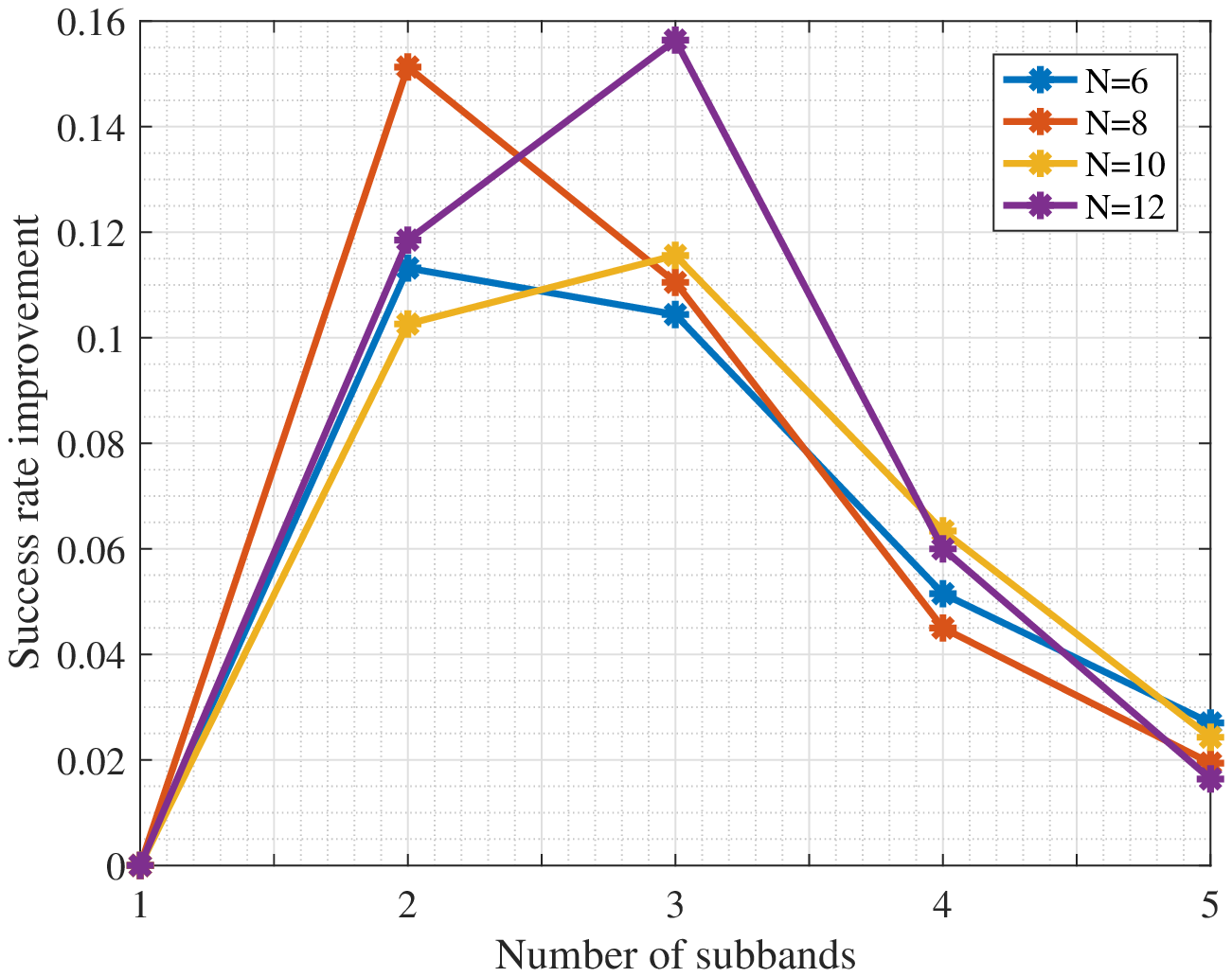}}
	\caption{Success rate improvement by the RL-based approach over the random and myopic policy versus the number of subbands. }
	\label{fig:raise_SR}
\end{figure*}

In Fig. \ref{fig:SR}, we compare the success rate achieved by the three approaches under different testing scenarios. Each scenario has a different combination of the number of cars and the number of subbands. In each subfigure, the number of cars is fixed and success rate versus the number of subbands is plotted. Generally, success rate achieved by the three approaches all increases with the number of subbands. The RL-based approach has the best performance in all scenarios and the random policy the worst.

In Fig. \ref{fig:raise_SR}, we plot the success rate improvement by the RL-based approach over the random (Fig. \ref{fig:raise_SR} (a)) and myopic policy ( Fig. \ref{fig:raise_SR} (b)) versus the number of subbands. In both subfigures, the success rate improvement first increases with the number of subbands and then decreases. Compared with the myopic policy, when there is only one subband, the two approaches equal; then, the improvement reaches its peak at 2 or 3 subbands, mostly exceeding $10\%$; when there are 5 subbands, the improvement drops below $3\%$. 
As the myopic policy is essentially to switch to a random subband when failure occurs, if there are 2 or 3 subbands, the probability of two radars switching to the same subbands is still large (the probability is $1/M$). However, the RL-based approach can use additional position and interference information, processed by the learned Q-networks, to avoid interference. The myopic policy catches up with the RL-based approach when the number of subbands increases because the probability of two radars switching to the same subbands becomes small. 
By analyzing the improvement of the RL-based approach over the myopic policy, it can be concluded that the RL-based approach is more advantageous when subbands are fewer. This means that using the proposed approach, we can divide the whole band into fewer subbands so that each subband can be allocated with more bandwidth for higher resolution.


Examining the relationship between the performance of the proposed approach and the number of subbands is instructive in dividing the spectrum. As is stated previously, the number of subbands is predefined majorly according to radar resolution or bandwidth requirement. Now, the interference avoidance performance can be another factor to be considered. More subbands guarantee higher success rate but result in lower range resolution due to bandwidth reduction. The proposed approach is verified to be more advantageous over the myopic policy with fewer subbands, which means that we can achieve the same interference avoidance performance with fewer subbands so that each subband can be assigned with more bandwidth for higher resolution. In other words, our approach makes it more effective to compromise between resolution and success rate.

\begin{figure*}
	\centering
	\subfloat[$N=10,M=3$]{\includegraphics[width=0.5\textwidth]{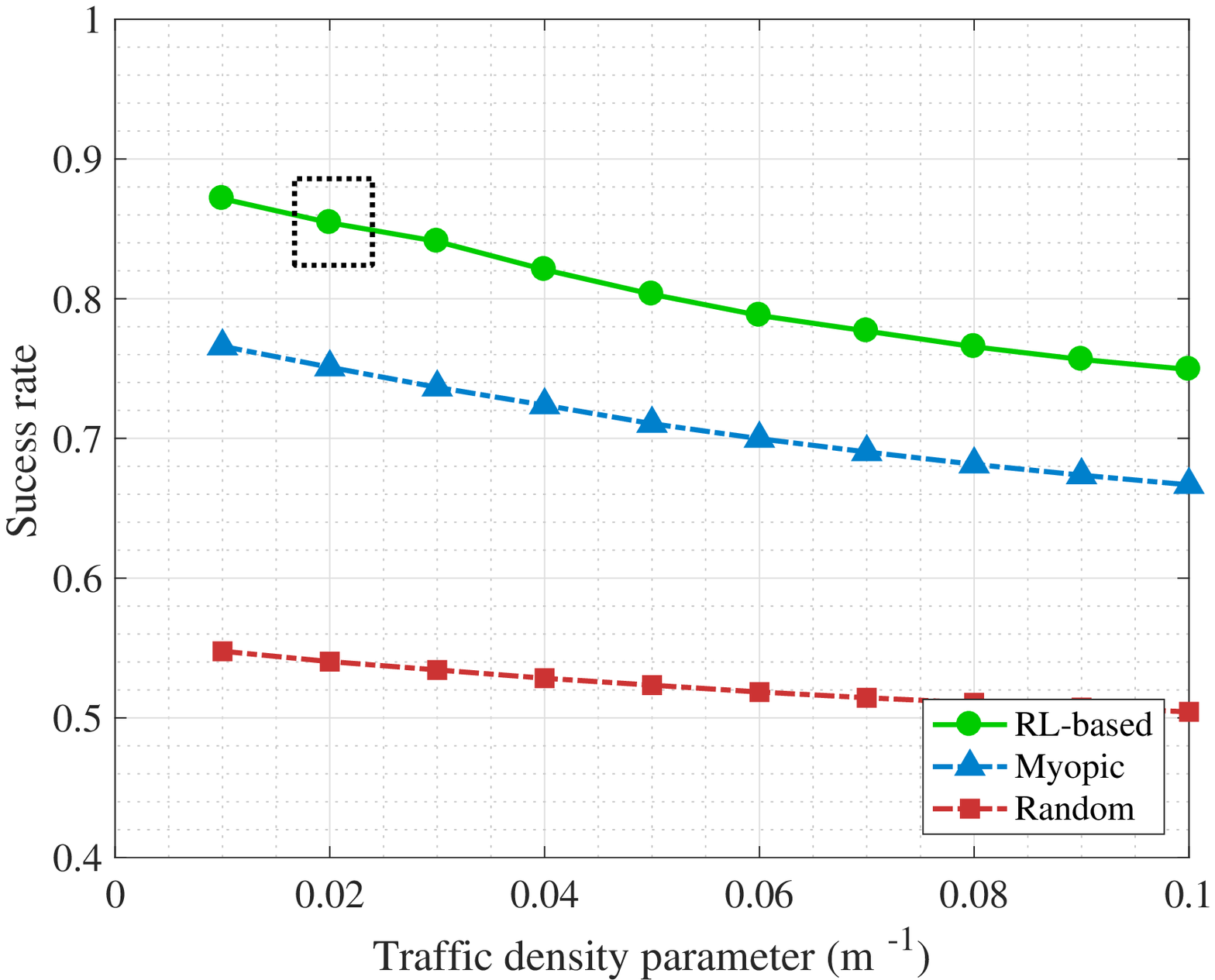}}
	\subfloat[$N=10,M=4$]{\includegraphics[width=0.5\textwidth]{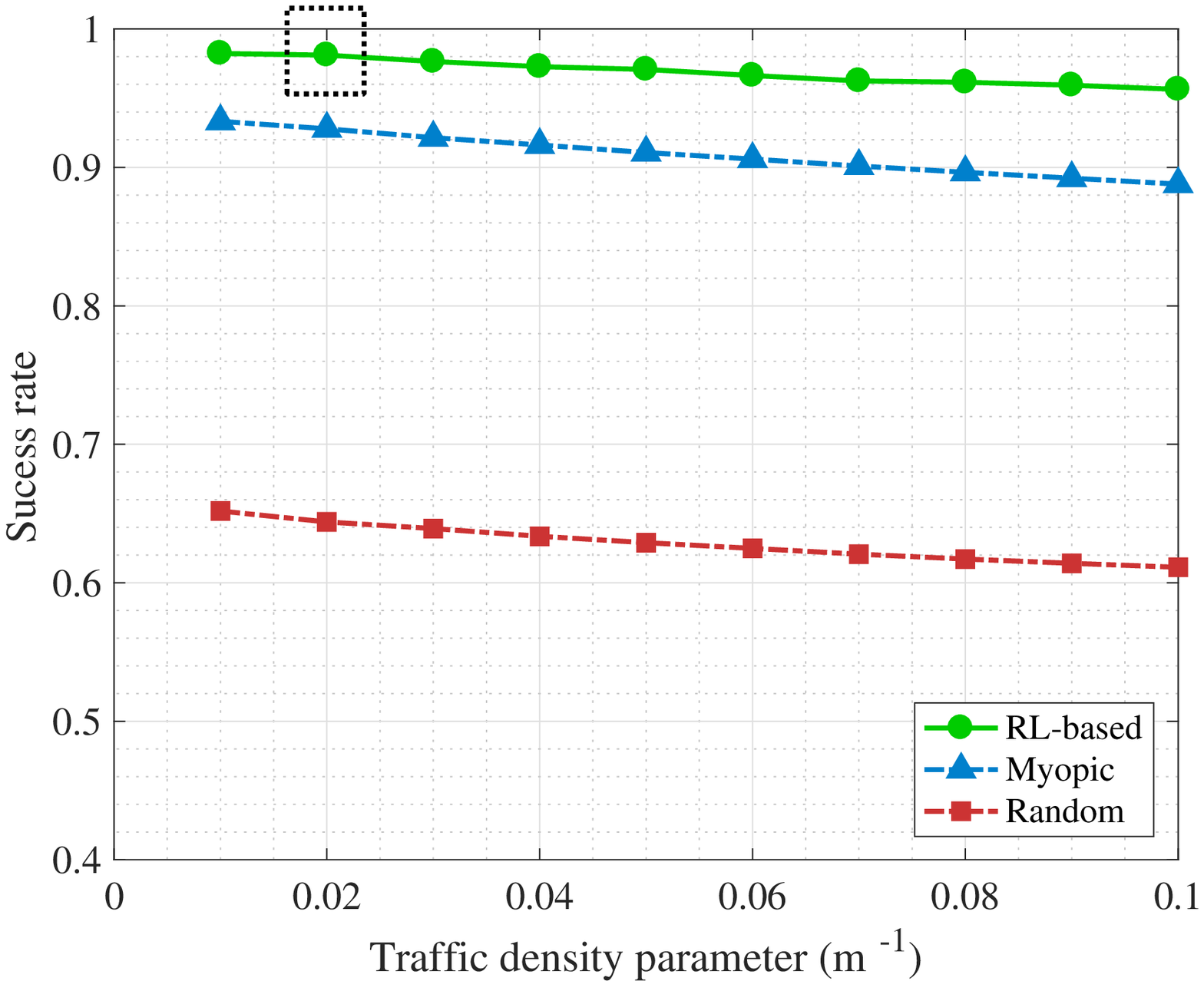}}\\
	\subfloat[$N=12,M=3$]{\includegraphics[width=0.5\textwidth]{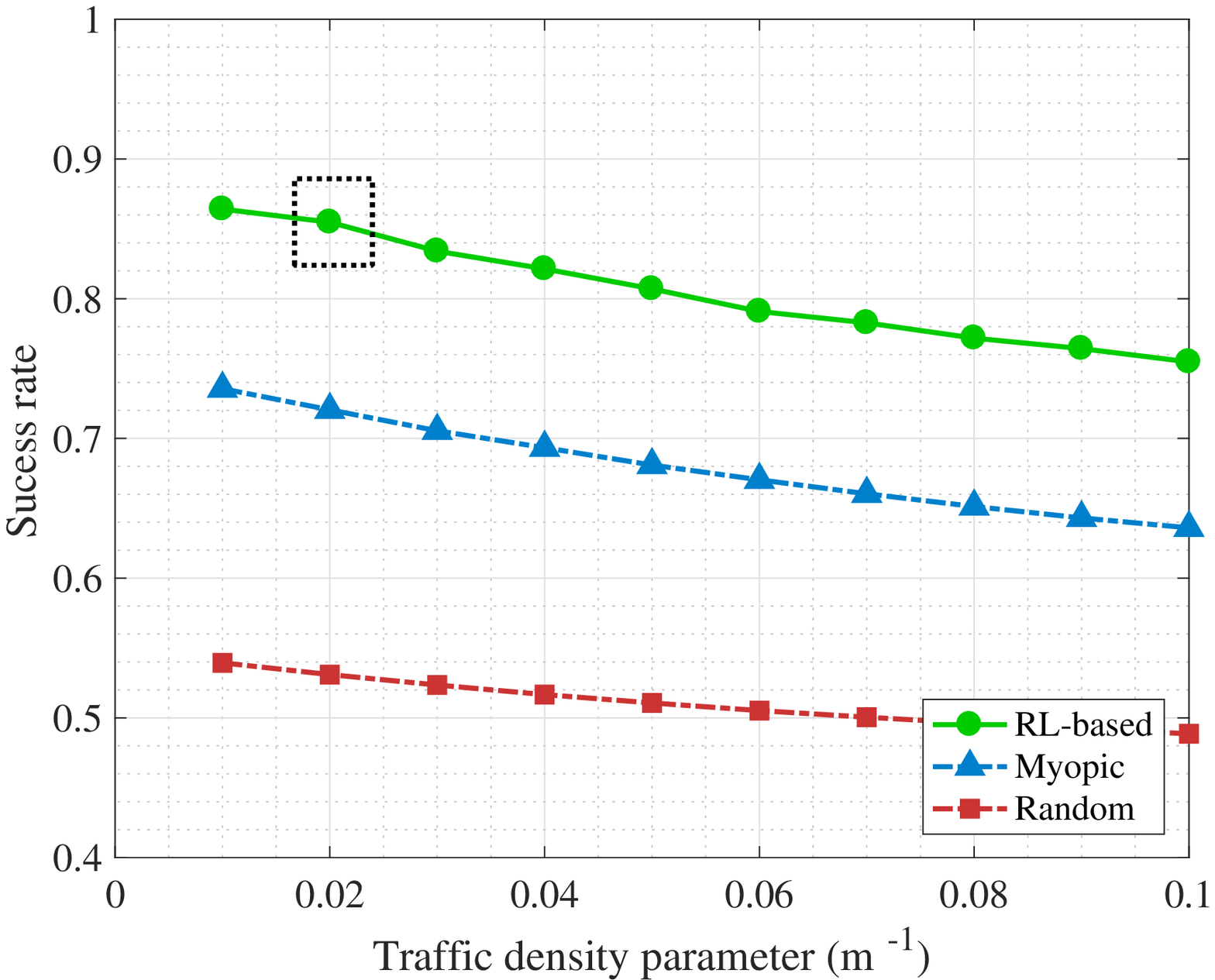}}
	\subfloat[$N=12,M=4$]{\includegraphics[width=0.5\textwidth]{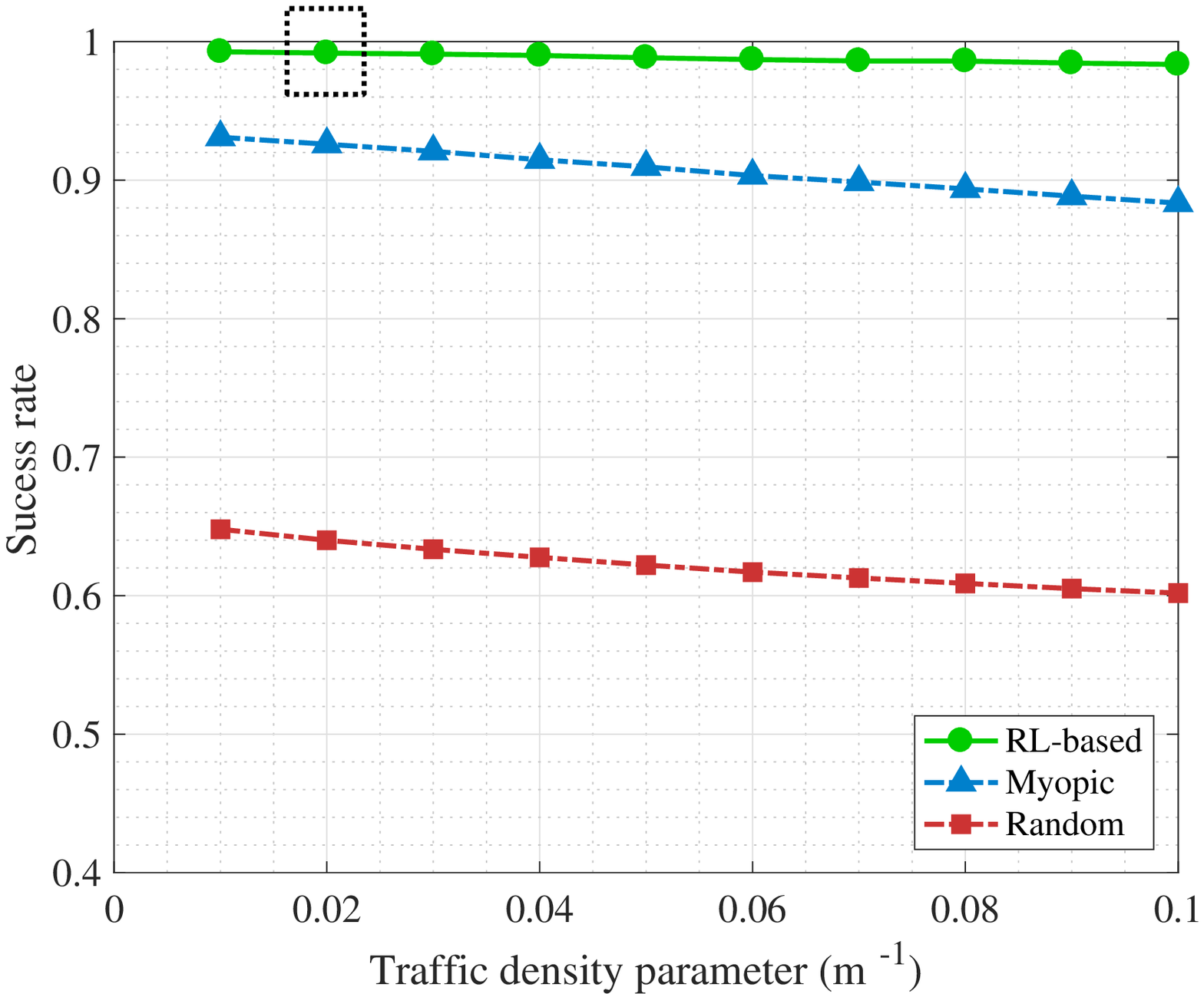}}
	\caption{Success rate achieved by the three approaches versus the traffic density parameter under different scenarios. The dashed box indicates that the Q-networks are trained when the traffic density parameter is $0.02$ and then the trained networks are  applied to other cases with different traffic density parameter.}
	\label{fig:SR_rho}
\end{figure*}

Next, we verify the robustness of the proposed approach in different road condition, such as the traffic density. In Fig. \ref{fig:SR_rho}, we plot success rate of the three approaches versus traffic density parameter $\rho$ under 4 different scenarios. 
In the RL-based approach, the Q-networks are trained when $ \rho=0.02$. Then the trained Q-networks are applied to other cases with different values of $\rho$ ranged from 0.01 to 0.1. 
Generally, for all three approaches, success rate decreases with the traffic density parameter increasing.
Compared to the myopic policy, the RL-based approach has a steady success rate improvement in the examined range of $\rho$, which shows that the trained network can be well generalized to other traffic density.

\begin{figure}[!t]
	\centering
	\includegraphics[width=0.5\textwidth]{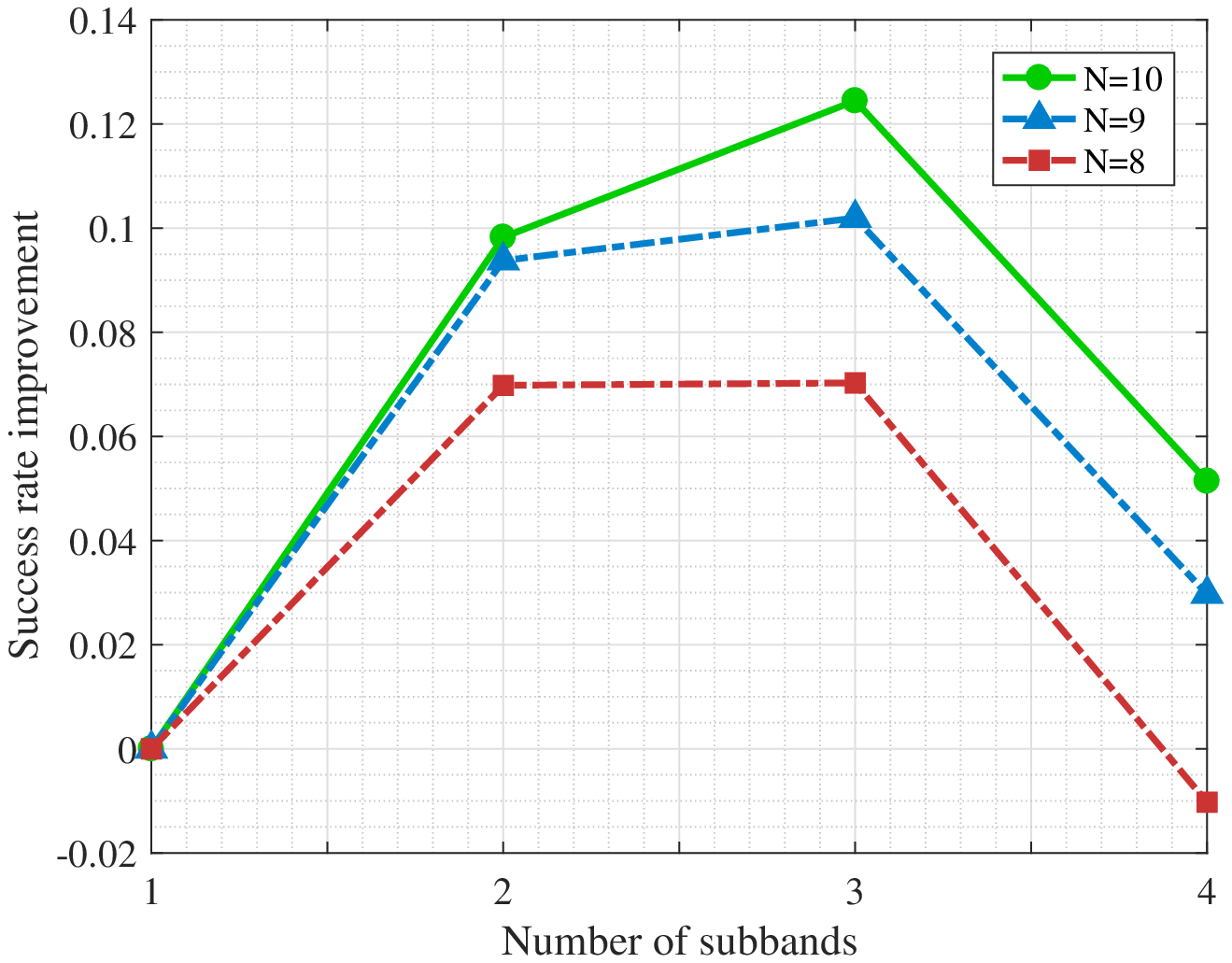}
	\caption{Success rate improvement over the myopic policy versus the number of subbands when applying the Q-networks trained for the 10-car scenario to the 10-, 9- and 8- car scenarios, respectively.}
	\label{fig:car_num_gen}
\end{figure}
\textcolor{black}{In the scenario model, it is assumed that the number of cars is fixed as $N$. As the Q-networks are trained offline, the pre-trained Q-networks for different number of cars can be stored in each car. Because the interfering power decreases with the distance in the inverse square law, we can consider a certain area within which cars interfere with each other but the interference caused by cars outside the area can be neglected. The number of cars can be acquired by each car via the popular vehicle-to-everything (V2X) communications, including vehicle-to-infrastructure communication \cite{chen2018vehicle}. Then, each car can select its own Q-network accordingly. The time between two adjacent communications can be much longer than $T$ because the number of cars in the area does not change fast. After the communication, each car can independently select subbands using the Q-networks.}

\textcolor{black}{Furthermore, we find it interesting that the Q-networks trained for $N_1$-car scenarios are also suitable for $N_2$-car ($N_2\le N_1$) scenarios. In the simulation, we apply the trained Q-networks for the 10-car scenario to the 9- and 8- car scenarios, respectively. Each of the 9 or 8 cars selects its Q-network out of the trained 10 Q-networks, which means the remaining 1 or 2 Q-networks are not used. In Fig. \ref{fig:car_num_gen}, we plot the success rate improvement over the myopic policy for each scenario. As Fig. \ref{fig:car_num_gen} shows, in the 9-car scenario, when the number of subbands is 2 or 3, the proposed approach can still gain around 10\% success rate improvement over the myopic policy. In the 8-car scenario, the improvement is around 6\%. As this simulation suggests, at the beginning, each car within an area can choose the Q-networks which are trained for a relative large number of cars. Then, new cars can just use the remaining Q-networks when they enter the area.}

\section{Conclusion}
\label{sec:conclusion}
In this paper, we study the interference avoiding problem for automotive radar using an RL-based decentralized spectrum allocation approach. With RL, each radar learns to choose a frequency subband merely according to its own observations with almost no communication. Considering a single  radar observation is inadequate, an LSTM neural network is incorporated in RL so that a subband is decided by combining both the present and past observations. Simulation experiments are conducted to verify the RL-based approach by comparing it with two commonly used spectrum allocation approaches, i.e., the random and myopic policy. It is shown that the RL-based approach gains a higher success rate improvement than the myopic policy with fewer subbands. Hence, the proposed approach makes it more effective to compromise between resolution and interference. 

The simulation model used in this paper is simplified to demonstrate the feasibility of the proposed approach. Future work will focus on constructing a simulation model which can better represent the much more complex road environments in the real world.

\appendices
\section{Derivation of the Beat Frequencies}
\label{app:A}
\textcolor{black}{
Recall that in (\ref{echo})(\ref{target_delay}), we give the target echo signal model. Let
\begin{equation}
\lambda = \frac{2v}{c}.
\end{equation}
The phase of the transmitted signal $s(t)$ and the target echo $e(t)$ are 
\begin{equation}
\phi_s(t)=\left\{\begin{array}{ll}
\dfrac{B}{2T_c}t^2+f_mt&0\le t<T_c\\
\\
\dfrac{B}{2T_c}\left(t-2T_c\right)^2+f_mt&T_c \le t < 2T_c \end{array}
\right.\\
\end{equation}
and
\begin{equation}
\phi_e(t)=\phi_s\left(\left(1+\lambda\right)t-\frac{2D}{c}\right),
\end{equation}
respectively. Then we calculate the instantaneous frequency of $s(t)$ and $e(t)$ for the up-chirp:
\begin{equation}
\frac{\partial \phi_s(t)}{\partial t}=\frac{B}{T_c}t+f_m
\end{equation}
and
\begin{equation}
\begin{split}
\frac{\partial \phi_e(t)}{\partial t}&=\frac{B}{T_c}\left((1+\lambda)t-\frac{2D}{c}\right)(1+\lambda)+f_m(1+\lambda)\\
&=\frac{\partial \phi_s(t)}{\partial t}+\frac{B}{T_c}(\lambda^2+2\lambda)t\\
&\quad-(1+\lambda)\frac{B}{T_c}\cdot\frac{2D}{c}+\lambda f_m,
\end{split}
\end{equation}
respectively. As $\lambda\ll 1,B\ll f_m$, we have
\begin{equation}
\frac{B}{T_c}(\lambda^2+2\lambda)t<  B(\lambda^2+2\lambda)\ll \lambda f_m.
\end{equation}
Therefore, the beat frequency of the up-chirp is 
\begin{equation}
f_b^\uparrow=\frac{\partial \phi_e(t)}{\partial t}- \frac{\partial \phi_s(t)}{\partial t}\approx-\frac{B}{T_c}\cdot\frac{2D}{c}+\frac{2v}{c}f_m.
\label{up_chirp_if}
\end{equation}
Likewise, for the down-chirp, the beat frequency is shown in (\ref{beat_freq2}). }

\begin{figure}[!t]
	\centering
	\includegraphics[width=0.5\textwidth]{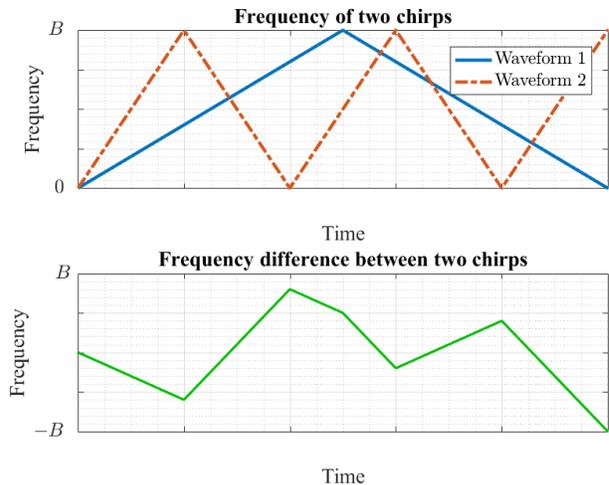}
	\caption{\textcolor{black}{Illustration showing that different chirp rates cause non-constant beat frequency.}}
	\label{fig:freq_diff}
\end{figure}
\section{Derivation of the Raise in the Noise Level}
\label{app:B}

\textcolor{black}{
If different radars uses different chirp rates, the frequency difference between two chirps is illustrated in Fig. \ref{fig:freq_diff}, which shows that the intermediate frequency signal, $r_{\text{IF}}(t)$, consists of several pieces of linear frequency modulation signals. Therefore, $r_{\text{IF}}(t)$ can be written as
\begin{equation}
r_{\text{IF}}(t)=\sum_{i=1}^{N_p}r_{p,i}(t),
\end{equation}
where $N_p$ is the number of the pieces of linear frequency modulation signals and $r_{p,i}(t)$ is the $i$th pieces of signal. The signal $r_{p,i}(t)$ is expressed as (\ref{r_pi}), in which $t_{i-1}, t_{i}$ are the starting and ending time of $r_{p,i}(t)$, $f_{i-1}, f_{i}$ are the starting and ending frequency, $k_i=(t_{i}-t_{i-1})/(f_{i}-f_{i-1})$ is the frequency modulation slope and $\phi_i$ is the initial phase.}

\begin{figure*}[t]
	\textcolor{black}{
		\begin{equation}
		\label{r_pi}
		r_{p,i}(t)=\textrm{rect}\left(\frac{t-t_{i-1}}{t_i-t_{i-1}}\right)\cdot\exp\left(j2\pi\left(\frac{1}{2}k_i\left(t-t_{i-1}\right)^2+f_{i-1}(t-t_{i-1})+\phi_i\right)\right)
		\end{equation}
		\begin{equation}
		R_{p,i}(f)=\mathcal{F}\left\{r_{p,i}(t)\right\}\approx \sqrt{\frac{\pi}{2k_i}}\cdot\textrm{rect}\left(\frac{f-f_{i-1}}{f_i-f_{i-1}}\right)\cdot\exp\left(j\frac{\pi}{4}+j\phi_i-j2\pi ft_{i-1}\right)
		\label{R_pi}
		\end{equation}}
\end{figure*}

\textcolor{black}{The Fourier transform of the $i$th piece of linear frequency modulation signal can be approximated by (\ref{R_pi}). Hence, we have
\begin{equation}
R(f)=\sum_{i=1}^{N_p} R_{p,i}(f).
\label{R_f2}
\end{equation}
As (\ref{R_pi}, \ref{R_f2}) indicates, the spectrum $R(f)$ can be approximately taken as the sum of several non-overlapping rectangular functions. Therefore, by using different chirp rate, ghost targets are avoided. Instead, the noise level is raised.}

%
%
%
%
%

\ifCLASSOPTIONcaptionsoff
  \newpage
\fi



%
\bibliographystyle{IEEEtran}

\bibliography{reference}

\newcounter{mytempeqncnt}

%

%






\end{document}